\documentclass[pra,aps,twocolumn,superscriptaddress,nofootinbib,showpacs]{revtex4}

\usepackage{latexsym}
\usepackage{hyperref}
\usepackage{amsmath}
\usepackage{xspace}
\usepackage{amssymb}
\usepackage{amsthm}

\newtheorem{theorem}{Theorem}
\newtheorem{lemma}{Lemma}
\newtheorem{proposition}{Proposition}

\newcommand{\sinc}{\mbox{sinc}}
\newcommand{\ad}{\mbox{ad}}

\begin{document}

\title{A geometric approach to quantum circuit lower bounds}
\author{Michael A. Nielsen} 
\thanks{nielsen@physics.uq.edu.au and www.qinfo.org/people/nielsen} 
\affiliation{School of Physical Sciences,
The University of Queensland, Brisbane, Queensland 4072, Australia}

\date{\today}

\begin{abstract}
  What is the minimal size quantum circuit required to exactly
  implement a specified $n$-qubit unitary operation, $U$, without the
  use of ancilla qubits?  We show that a lower bound on the minimal
  size is provided by the length of the minimal geodesic between $U$
  and the identity, $I$, where length is defined by a suitable Finsler
  metric on the manifold $SU(2^n)$.  The geodesic curves on these
  manifolds have the striking property that once an initial position
  and velocity are set, the remainder of the geodesic is completely
  determined by a second order differential equation known as the
  geodesic equation.  This is in contrast with the usual case in
  circuit design, either classical or quantum, where being given part
  of an optimal circuit does not obviously assist in the design of the
  rest of the circuit.  Geodesic analysis thus offers a potentially
  powerful approach to the problem of proving quantum circuit lower
  bounds.  In this paper we construct several Finsler metrics whose
  minimal length geodesics provide lower bounds on quantum circuit
  size.  For each Finsler metric we give a procedure to compute the
  corresponding geodesic equation.  We also construct a large class of
  solutions to the geodesic equation, which we call \emph{Pauli
    geodesics}, since they arise from isometries generated by the
  Pauli group.  For any unitary $U$ diagonal in the computational
  basis, we show that: (a) provided the minimal length geodesic is
  unique, it must be a Pauli geodesic; (b) finding the length of the
  minimal Pauli geodesic passing from $I$ to $U$ is equivalent to
  solving an exponential size instance of the closest vector in a
  lattice problem (CVP); and (c) all but a doubly exponentially small
  fraction of such unitaries have minimal Pauli geodesics of
  exponential length.
\end{abstract}



\maketitle

\section{Introduction}
\label{sec:intro}

\subsection{Overview}

A central problem of quantum computation is to determine the most
efficient way of implementing a desired unitary operation.  Although
insight into this problem has been obtained for certain specific
unitary operations, no useful general techniques for determining the
most efficient implementation are known.

The interest in this problem arises from the desire to find classes of
unitary operations which can be implemented \emph{efficiently}, i.e.,
using polynomial resources.  Using a non-constructive
measure-theoretic argument, Knill~\cite{Knill95a} has shown that a
generic unitary operation requires exponentially many quantum gates
even to approximate.  Despite this result, no explicit construction of
a natural family of unitary operations requiring exponential size
quantum circuits is known.

An analogous situation holds classically, where
Shannon~\cite{Shannon49a} (see Theorem~4.3 on page~82
of~\cite{Papadimitriou94a}) used a non-constructive counting argument
to show that most Boolean functions $f : \{0,1\}^n \rightarrow \{ 0,
1\}$ require circuits of exponential size to compute.  Despite this
result, no explicit construction of a natural family of functions
requiring exponential size circuits is known.

The lack of explicit constructions of hard-to-compute operations is
symptomatic of the general difficulty encountered in proving lower
bounds on the computational resources required to synthesize specified
classes of operations, both quantum and classical.  The most
celebrated instance of this difficulty is, of course, the problem of
proving $\mathbf{P} \neq \mathbf{NP}$.  More generally, computer
scientists suspect many separations between computational complexity
classes, but techniques to prove them are elusive.

The problem motivating the present paper is inspired by the problems
just described, but is more restricted in scope.  Suppose $U$ is a
special unitary operation on $n$ qubits, i.e., a $2^n \times 2^n$
unitary operation with unit determinant\footnote{In this paper we work
  mainly with $SU(2^n)$, rather than $U(2^n)$.  As a consequence, all
  unitaries are assumed to have unit determinant, unless otherwise
  remarked.}.  Let ${\cal G}$ be a set of unitary gates which is
universal on $n$ qubits, e.g., the set of single-qubit unitary
operations and any fixed entangling two-qubit
gate~\cite{Brylinski02a,Bremner02a,Zhang03a}.  We require ${\cal G}$
to be \emph{exactly} universal, i.e., the group generated by ${\cal
  G}$ should be $SU(2^n)$, not some dense subset.  Then we define
$m_{\cal G}(U)$ to be the minimal number of gates from ${\cal G}$
required to exactly synthesize $U$.

In this paper we explain how to introduce a metric $d(\cdot,\cdot)$ on
$SU(2^n)$ such that $d(I,U) \leq m_{\cal G}(U)$, where $I$ is the
$n$-qubit identity operation.  Thus the metric $d$ provides a
\emph{lower bound} on the number of gates required to implement $U$.
We define our metric by first specifying a structure known as a
\emph{local metric}, which can be thought of as assigning a distance
to points nearby on the manifold.  This local metric induces a natural
notion of curve length, which can then be used to define
$d(\cdot,\cdot)$ as the infimum over lengths of curves between two
points.

\subsection{Motivating ideas}
\label{subsec:motivating_ideas}

Two key ideas motivate our geometric approach to the problem of
proving lower bounds on $m_{\cal G}(U)$.  The first idea is that it is
easier to minimize a smooth function on a smooth space rather than a
general function on a discrete space, and thus whenever possible we
replace discrete structures by smooth structures\footnote{In a related
  vein, Linial~\cite{Linial02a} has written a stimulating survey
  devoted to the use of geometric ideas in combinatorics.}.  This is,
of course, an idea familiar to any undergraduate: it is far easier to
minimize a smooth function defined on the reals than it is to minimize
that function when restricted to the integers.  The reason, of course,
is that the differential calculus enables us to minimize smooth
functions on smooth spaces, using the powerful principle that such a
function $f(\cdot)$ should be stationary at a minimum, and thus
satisfy the equation $f'(x) = 0$.

This observation motivates us to reformulate the problem of finding a
minimal quantum circuit --- essentially, an optimization over the
discrete space of possible quantum circuits\footnote{Of course, some
  (though not all) universal gate sets have a smooth structure.  By
  ``discrete'' we mean here that the application of a quantum gate is
  a discrete event, and so the number of quantum gates applied is
  necessarily a non-negative integer.} --- with a closely related
problem, which we call the \emph{Hamiltonian control problem}.  In the
Hamiltonian control problem we attempt to minimize a smooth cost
function with respect to a smooth set of Hamiltonian control
functions.  This formulation allows us to apply the principle that a
smooth functional is stationary at its minimum.  Technically, we carry
this out by using the calculus of variations to study the minimal cost
Hamiltonian control function.  As we describe below, this idea leads
naturally to a geodesic equation whose solutions are control functions
which are local minima of the cost function.

A second key idea, overlapping the first, motivates our approach.
This idea is built on an analogy to the many principles of physics
which can be formulated in one of two equivalent ways: a description
in which the motion of a particle is described in terms of a local
force law, and a description in which motion is described in terms of
minimizing some globally defined functional.  To make this analogy
precise we use the example of test particle motion in general
relativity, but similar remarks may be made in many other areas of
physics, including classical mechanics (Newtonian versus Lagrangian
formulations), and optics (geometric optics versus Fermat's
principle).
  
It is a basic principle of general relativity that test particles move
along geodesics of spacetime, i.e., move so as to minimize a globally
defined functional, the pseudo-Riemannian distance.  This principle
turns out to be equivalent to the particle following the geodesic
equation,
\begin{eqnarray} \label{eq:geodesic}
  \frac{d^2 x^j}{dt^2} + \Gamma^{j}_{kl} \frac{dx^k}{dt} \frac{dx^l}{dt} = 0,
\end{eqnarray}
where $x^j$ are co-ordinates for the position on the manifold, we sum
over the repeated indices $k$ and $l$, and the Christoffel symbols
$\Gamma^j_{kl}$ are real numbers determined by the local geometry of
spacetime.  This reformulation shows that minimizing the length
traversed is equivalent to following what is essentially a local force
law: the geodesic equation tells us how a particle ought to
accelerate, given its current velocity and the local geometry.  More
generally, on any Riemannian or pseudo-Riemannian manifold the problem
of geodesic motion turns out to be equivalent to following a local
force law, the same geodesic equation of Equation~(\ref{eq:geodesic}).

This situation is in sharp contrast to the problem of finding an
optimal circuit to compute a function.  Suppose someone gives us a
partially complete circuit to compute a function, $f$, and asks us to
complete the circuit.  In general, there are no useful techniques for
determining the best way of completing the circuit, short of an
exhaustive search.  But given an arbitrarily small arc along a
geodesic on a Riemannian manifold, the remainder of the geodesic is
completely determined by the geodesic equation.  Indeed, provided we
know the velocity at some given point the remainder of the geodesic is
completely determined by the geodesic equation.

This analogy motivates our formulating the Hamiltonian control problem
so that the cost function to be minimized is a local metric structure
on a suitable type of manifold, which we shall argue below is a
\emph{Finsler} manifold, a type of manifold generalizing the
Riemannian manifolds most familiar to physicists.  Just as for
Riemannian manifolds, we will see that the geodesics on a Finsler
manifold are determined by a geodesic equation of the form of
Equation~(\ref{eq:geodesic}), but where the coefficients
$\Gamma^{j}_{kl}$ are a generalized type of Christoffel symbol for the
Finsler manifold.  Thus, once the initial position and velocity (or
any small arc) are known, the remainder of the geodesic is uniquely
determined by the geodesic equation.

An important caveat to this otherwise encouraging situation is that
while the solutions to the geodesic equation are local minima of the
cost function, they may not be global minima\footnote{The analogous
  situation in ordinary calculus is, of course, the fact that $f'(x) =
  0$ may have multiple solutions.}.  That is, there may be multiple
geodesics connecting $I$ and $U$, and we will see some explicit
examples of this later.  Nonetheless, the minimal length curve is
guaranteed to be a geodesic\footnote{For this reason, we use the terms
  ``minimal length curve'' and ``minimal length geodesic''
  interchangeably.}, and thus geodesic analysis offers a potentially
powerful approach to proving lower bounds on $m_{\cal G}(U)$.

\subsection{Structure and main results}

\textbf{The local metric is a right-invariant ${\cal G}$-bounding
  Finsler metric:} As just described, our strategy is to define and
study suitable classes of local metrics on the manifold $SU(2^n)$.
What properties should these local metrics have?  In order to find a
suitable local metric our strategy is to sequentially impose more and
more restrictive conditions, motivated by various properties that we
desire the minimal length curves to have.

We begin in Section~\ref{sec:metrics_on_manifolds}
(Subsections~\ref{subsec:HCP}
and~\ref{subsec:cost-function-and-metrics}), with a simple general
argument motivating the use of local metrics as a measure of the cost
of implementing a unitary operation.  In
Section~\ref{sec:metrics_on_manifolds}
(Subsection~\ref{subsec:cost-function-and-metrics}) a simple extension
of the same argument is used to motivate the condition that the local
metric be right-invariant, which corresponds to the physical
requirement that the cost of applying a particular Hamiltonian should
not depend on when that Hamiltonian is applied, i.e., it is an
expression of homogeneity.

In Section~\ref{sec:metrics_on_manifolds}
(Subsection~\ref{subsec:minimal_curves}) we impose the additional
requirement that the local metric should be capable of providing lower
bounds on gate complexity, i.e., $d_F(I,U) \leq m_{\cal G}(U)$, where
$d_F(I,U)$ is the distance between $I$ and $U$ induced by the local
metric, which we denote by $F$.  In particular, we prove a simple
theorem, giving sufficient conditions on $F$ in order that the
inequality $d_F(I,U) \leq m_{\cal G}(U)$ hold.  We call local metrics
satisfying this condition \emph{${\cal G}$-bounding}.

Finally, in Section~\ref{sec:metrics_on_manifolds}
(Subsection~\ref{subsec:finsler-definition}) we impose the additional
requirement that the local metric have sufficient smoothness and
convexity properties to allow the calculus of variations to be applied
to study the minimal length curves.  We will show that this is
equivalent to requiring that the local metric be a Finsler metric.

Finsler metrics are a class of local metrics generalizing the
Riemannian metrics familiar to physicists from the study of general
relativity.  In Riemannian geometry the length of a small displacement
on the manifold is determined by the square root of some quadratic
form in the displacement.  On a Finsler manifold, this special form
for the local metric is replaced by a general norm function, subject
only to the most general smoothness and convexity properties
sufficient to ensure that a second order differential equation holds
for the geodesics.  Essentially, Finsler metrics may be viewed as the
most general class of local metrics giving rise to such a geodesic
equation.

Summing up, the main result of Section~\ref{sec:metrics_on_manifolds}
(Subsections~\ref{subsec:HCP} through~\ref{subsec:finsler-definition})
is that the most suitable local metric structure is a right-invariant
${\cal G}$-bounding Finsler metric.

\textbf{Construction of local metrics providing lower bounds on
  $m_{\cal G}(U)$:} In Section~\ref{sec:metrics_on_manifolds}
(Subsection~\ref{subsec:examples}) we construct three important
families of right-invariant ${\cal G}$-bounding local metrics, which
we denote $F_1, F_p$ and $F_q$.  As each of these local metrics is
${\cal G}$-bounding, they all give rise to lower bounds on $m_{\cal
  G}(U)$, through the results of
Subsection~\ref{subsec:minimal_curves}.  However, $F_1$ and $F_p$ lack
some of the smoothness and convexity properties required by Finsler
metrics.  In order to analyse $F_1$ and $F_p$ within the desired
framework of Finsler geometry, in
Appendix~\ref{app:approximating-metrics} we construct a parameterized
family of right-invariant Finsler metrics $F_{1\Delta}$ and
$F_{p\Delta}$, with the property that $F_{1\Delta} \rightarrow F_1$
and $F_{p\Delta} \rightarrow F_p$ as $\Delta \rightarrow 0$.  That is,
we can approximate $F_1$ and $F_p$ as well as desired using suitable
families of right-invariant Finsler metrics, and thus study them using
the geodesic equation of Finsler geometry.

\textbf{Computing the geodesic equation:} In
Section~\ref{sec:geodesic_equation} we explain how to compute the
geodesic equation for each of our families of right-invariant Finsler
metrics $F_{1\Delta}, F_{p\Delta}$, and $F_q$.  The main tool used in
the computation of the geodesic equation is a generalization of the
Baker-Campbell-Hausdorff formula used by physicists, which is used to
accomplish a necessary change of co-ordinates on $SU(2^n)$.  As a
simple illustration of the utility of the geodesic equation, in
Section~\ref{sec:geodesic_equation}
(Subsection~\ref{subsec:direct_sum}) we consider the effect
ancilla qubits have on the minimal length curves, showing that for
Finsler metrics satisfying suitable conditions --- $F_q$ is an example
of such a Finsler metric --- there is a neighbourhood of the identity
in which the presence or absence of ancilla qubits does not affect the
minimal length curves.
  
\textbf{Construction of the Pauli geodesics:} In
Section~\ref{sec:geodesic_properties}
(Subsections~\ref{subsec:symmetries} and~\ref{subsec:Pauli}) we
construct a class of curves in $SU(2^n)$ which are geodesics for all
three of the Finsler metrics $F_{1\Delta}, F_{p\Delta}$, and $F_q$.
We call these \emph{Pauli geodesics}, as they arise naturally from a
class of isometries associated with the Pauli group.

\textbf{Finding minimal length Pauli geodesics is equivalent to
  solving an instance of closest vector in a lattice:} In general,
many geodesics may connect any two points on $SU(2^n)$, and the
problem of finding the minimal length curve connecting two points may
be viewed as the problem of finding the minimal length geodesic.  We
show in Section~\ref{sec:geodesic_properties}
(Subsection~\ref{subsec:CVP}) that the problem of finding the minimal
length \emph{Pauli} geodesic through a unitary $U$ which is diagonal
in the computational basis is equivalent to solving an exponential
size instance of the closest vector in a lattice problem (CVP), well
known from computer science.  

This reduction to CVP is perhaps somewhat ironic, given our general
philosophy of replacing discrete structures by smooth structures.
However, we will see that the CVP instance is far simpler and has a
much more elegant structure than the original problem of finding a
minimal-size quantum circuit.

The reduction to CVP is only for the problem of finding the minimal
length Pauli geodesic, not the minimal geodesic of any type.  Also in
Subsection~\ref{subsec:CVP}, we show that provided there is a
\emph{unique} minimal length geodesic through $U$, then the minimal
length geodesic \emph{is} a Pauli geodesic, and thus the solution of
the CVP instance will give the distance $d_F(I,U)$, and so provide a
lower bound on $m_{\cal G}(U)$.  Unfortunately, we are not able to say
how generically this situation holds.  Standard examples of Riemannian
manifolds such as the sphere, flat space and the hyperbolic spaces
suggest it may be true of many or perhaps most $U$.  Against this, in
Subsection~\ref{subsec:discussion-of-Pauli-geodesics} we give an
example where the minimal Pauli geodesic is provably \emph{not} the
minimal length geodesic of any type.

\textbf{Minimal length Pauli geodesics of exponential length exist:}
Using the connection to CVP, in Section~\ref{sec:geodesic_properties}
(in Subsection~\ref{subsec:vol-argument}) we prove that the
overwhelming majority of unitaries diagonal in the computational basis
have minimal length Pauli geodesics of exponential length.  The method
of proof is a volume argument, suggested to the author by Oded~Regev.

\textbf{Caveats:} Several additional caveats to our results should be
made clear.

The first caveat is that although we have made considerable progress
understanding the geodesic structure of our local metrics, we are a
long way from a complete understanding of either the geodesics or the
minimal length curves of those local metrics.  As an example of the
type of basic question that is still unresolved, we do not even know
for sure if minimal curves of exponential length exist.

These difficulties are perhaps not surprising, as in general it is an
extremely difficult problem to understand the minimal length geodesics
on a manifold.  Indeed, there are few manifolds even of Riemannian
type for which the geodesics are completely understood.  This paper
should thus be viewed as a first step toward an understanding of these
minimal length geodesics.

The second caveat is that although we show that $d_F(I,U) \leq m_{\cal
  G}(U)$ for right-invariant ${\cal G}$-bounding Finsler metrics, $F$,
it is by no means clear how to choose $F$ in such a way as to achieve
the desirable property that $m_{\cal G}(U)$ and $d_F(I,U)$ be
polynomially equivalent.  I conjecture that all three of $F_{1\Delta},
F_{p\Delta}$, and $F_q$ have this property, for suitable parameter
choices.  The one concrete step in this direction we take is to show
that as $\Delta \rightarrow 0$ the distance $d_{F_{1\Delta}}(I,U)$ can
be interpreted as the minimal time required to generate $U$, using a
set of control Hamiltonians each of which can be efficiently simulated
in the standard quantum circuit model.

The third caveat is a reiteration and extension of the earlier point
that while understanding the behaviour of $m_{\cal G}(U)$ would be
extremely interesting in its own right, it is really a toy version of
some much more interesting problems in quantum computational
complexity. There are three important ways the determination of
$m_{\cal G}(U)$ falls short of the problems of interest in quantum
computational complexity: (1) the requirement that the synthesis of
$U$ be exact, rather than approximate; (2) the requirement that this
synthesis be performed without the benefit of additional workspace,
i.e., without ancilla qubits initially prepared in a standard state;
and (3) the lack of a uniformity requirement on the circuit
implementing $U$, i.e., there is no requirement that there be a
polynomial-time Turing machine efficiently generating a description of
the circuit. Obviously it is to be hoped that these shortcomings can
be mitigated by future extensions of the present approach.

The fourth caveat is in relation to the presentation of the paper. The
paper is intended to be accessible to physicists, mathematicians, and
computer scientists, especially those involved in quantum information
science, but makes use of ideas from certain areas of mathematics ---
differential geometry, Finsler geometry, and the vectorization of
matrix equations --- that may be unfamiliar to many readers.  As the
goal of the paper is primarily to synthesize a program for
investigating quantum lower bounds, rather than to solve specific
technical problems previously considered inaccessible, I have included
considerable introductory material and references, as well as
describing some arguments in considerable detail, in order to make the
broad picture as clear as possible.

\subsection{Prior work}
\label{subsec:prior-work}

We divide prior work up into research from four different points of
view: optimal quantum control, universality constructions for quantum
circuits, quantum circuit design, and computational complexity theory.

\textbf{Optimal quantum control.} The work most similar in spirit to
the present paper comes from the field of quantum control,
particularly optimal quantum control.  Quantum control is a large
field, and we will not attempt to comprehensively survey it here ---
see, e.g.,~\cite{Gordon97a,Rabitz00a,Rice00a,Shapiro03a} for an entry
into the literature, and further references.  

Only a relatively small part of the quantum control literature has
been concerned with time-optimal methods for generating unitary
operations.  These methods may be subdivided into two (overlapping)
approaches: those based on geometric control theory, and those based
on using the calculus of variations to minimize some global cost
functional without a direct geometric interpretation.

The research most closely related to the present paper is the work on
geometric quantum control pursued by Khaneja, Brockett and
Glaser~\cite{Khaneja01a} (see also~\cite{Khaneja01b,Khaneja02a}), by
Zhang and Whaley~\cite{Zhang04a}, and by Dirr \emph{et
  al}~\cite{Dirr04a}.  We now briefly outline the approach taken in
this prior work, in order to contrast it with the approach taken in
the present paper.

These prior works formulate the problem of quantum control as the
problem of synthesizing a unitary operation $U$ using a time-dependent
control Hamiltonian $H = H_d + \sum_j v_j H_j$, where $H_j$ are
control Hamiltonians, $v_j$ are real control functions, and $H_d$ is
the drift Hamiltonian.  The goal is to synthesize $U$ in the minimal
possible time.  It is assumed that the control functions $v_j$ can be
made arbitrarily intense, for no cost, but the Lie group $K$ which
they generate is a strict subgroup of the total Lie group $SU(2^n)$.
$K$ might be, for example, the space of local unitary operations on
$n$ qubits, while $H_d$ is some global entangling Hamiltonian
connecting all the qubits~\cite{Dodd02a}.  Thus, the time taken to
synthesize $U$ is just the total time for which the drift Hamiltonian
$H_d$ is applied.

Khaneja \emph{et al}~\cite{Khaneja01a} show that this problem is
equivalent to finding the minimal length geodesics on the coset space
$SU(2^n) / K$, when that space is equipped with a suitable metric
structure.  Furthermore, they show that in the special case when
$SU(2^n)/K$ is a Riemannian symmetric space these geodesics have an
exceptionally simply structure that enable the minimal time to be
calculated exactly.  This is the case, for example, for $SU(4) / SU(2)
\otimes SU(2)$.

The power of this approach comes from the connection to the theory of
symmetric spaces, which have a beautiful theory that is exceptionally
well understood (see, e.g.,~\cite{Helgason01a}).  This is also its
limitation, for it is only in very special cases that $SU(2^n) / K$ is
a symmetric space.  See, e.g.,~\cite{Khaneja01b,Khaneja02a} for a
discussion of the limitations on $K$ imposed by this requirement.  In
practice, Khaneja \emph{et al}, Zhang and Whaley, and Dirr~\emph{et
  al} were all limited to studying geodesics for special cases where
$n \leq 3$.  (We note that some closely related results for $n=2$ have
been obtained in~\cite{Vidal02a,Hammerer02a,Childs03d}, using a
different approach based on the theory of majorization.)

Our approach is similar in spirit to this prior work, but differs
significantly in substance.  We do not identify any special subgroup
$K$, and thus work directly with the space $SU(2^n)$, using the
general framework of Finsler geometry, rather than the Riemannian
geometry of $SU(2^n)/K$.  In this framework, we relate the length of
the minimal Finsler geodesic to the minimal size quantum circuit.  As
our interest is motivated by quantum computation, we are primarily
interested in the case of an arbitrary number of qubits, $n$, and we
succeed in constructing geodesics valid for any $n$, and obtaining
some general (albeit, limited) results about the minimal length
geodesics for arbitrary $n$.

Optimal control theory has also given rise to a second strand of work
related to the present paper, with a rather more extensive literature
than the quantum geometric control literature.  Rather than reviewing
all this literature in detail, we refer the reader to a recent
sample~\cite{Rangan01a,DAlessandro01a,Tesch02a,Palao02a,Palao03a,Sklarz04a},
and the references therein.

Broadly speaking, the typical setting for this work is the problem of
finding the optimal way of generating a one- or two-qubit quantum
gate, using a specified Hamiltonian (e.g., a two-level atom coupled to
an external electromagnetic field) containing one or more control
parameters.  An \emph{ad hoc} functional is constructed, representing
the cost of generating the gate in terms of quantities such as the
power consumed.  The calculus of variations is then employed to derive
a condition for that functional to be maximized, typically resulting
in a two-point boundary value problem for some second order
differential equation, which is then solved numerically using
iterative techniques.  This body of work is thus much more concerned
with obtaining numerical results for specific Hamiltonians and
specific one- and two-body unitaries, rather than the general
$n$-qubit questions of most interest to us.

\textbf{Universality constructions for quantum circuits.} Researchers
working on universality constructions for quantum circuits have done
considerable work optimizing their constructions.  This began in the
early papers by Barenco \emph{et al}~\cite{Barenco95a} and
Knill~\cite{Knill95a}, who showed that the universality constructions
in~\cite{Barenco95a} are near-optimal for a \emph{generic} unitary
operation.  This work has subsequently been improved by many groups;
see, for example,~\cite{Tucci99a,Cybenko01a,Shende04a,Bergholm04a},
and references therein.  This line of investigation appears
superficially to be closely related to the topic of the present paper,
but that appearance is misleading.  The reason is that this prior work
investigates constructions which are only \emph{generically} optimal,
and there is no reason to believe that the constructions obtained in
any of these papers will be optimal for any \emph{particular} unitary
operation\footnote{A notable exception is that the algorithm
  in~\cite{Tucci99a} does reproduce the fast quantum Fourier transform
  circuit.  However,~\cite{Tucci99a} notes that the construction in
  that paper is not optimal in general.}, and thus they cannot be used
to deduce lower bounds on the minimal number of circuit elements
required to synthesize a specific unitary operation.

\textbf{Optimal quantum circuit design.} Another topic which has
attracted considerable prior interest is the design of optimal quantum
circuits for specific tasks.  This has become a major topic of ongoing
investigation; unfortunately no general survey exists, and a list of
references would run to many hundreds.  However, the key point is that
these papers derive optimal or near-optimal circuits only for certain
special classes of unitary operations, e.g., Cleve and
Watrous'~\cite{Cleve00a} fast parallel circuits for the quantum
Fourier transform.  Thus this work does not provide a general approach
to the problem of finding optimal circuits for unitary operations, nor
for the problem of finding lower bounds on the number of quantum gates
required to perform a given (but arbitrary) unitary operation.

\textbf{Computational complexity.} The theory of computational
complexity exists in large part, of course, to analyse the time cost
of computation, in both classical and quantum computing models.
General references are~\cite{Papadimitriou94a,Sipser97a}.  

Within quantum computational complexity, the work of most relevance to
the present paper is the work on oracle lower bounds, a selection of
which may be found in~\cite{Bennett97b,Beals98a,Ambainis02a}; see also
the references therein.  The oracle setting offers substantial
technical simplifications when compared with the problem of proving
unconditional lower bounds on the difficulty of synthesizing unitary
operations, but it is also widely regarded as a much less interesting
setting.  The results in the present paper are much less complete than
some of the results obtained in the oracle setting, but have the
advantage of being in the unconditional setting.

Within classical computational complexity, it is worth noting a
surface resemblance between the present work and the approach to the
$\mathbf{P} \neq \mathbf{NP}$ problem due to Mulmuley and
Sahoni~\cite{Mulmuley02a,Mulmuley01a,Mulmuley05a} (see
also~\cite{Regan02a}).  This work also uses geometric techniques to
address the problem of proving lower bounds.  However, the techniques
used are from algebraic geometry, based on geometric invariant theory,
and thus are not obviously related to the ideas used in the present
paper, which are based on Riemann and Finsler geometry.

\section{The Hamiltonian control problem and metrics on manifolds}
\label{sec:metrics_on_manifolds}

%
%
In this section we introduce the \emph{Hamiltonian control problem},
whose goal is to find a time-dependent Hamiltonian $H(t)$ synthesizing
$U$.  For a given Hamiltonian we then define a corresponding
\emph{cost}, which is a functional of $H(\cdot)$.  We argue on general
grounds that the cost function ought to arise from a right-invariant
Finsler metric on the manifold $SU(2^n)$, and provide general
conditions in order that such a cost function provide a lower bound on
$m_{\cal G}(U)$.  Furthermore, we introduce several Finsler metrics
satisfying these conditions, and discuss how effective each of these
Finsler metrics is likely to be as a means of proving quantum circuit
lower bounds.

%
%
\subsection{The Hamiltonian control problem} 
\label{subsec:HCP}

Let $U$ be a special unitary operation on $n$ qubits.  Our goal is to
synthesize $U$ using a traceless\footnote{\emph{A priori} there is no
  need for the Hamiltonian to be traceless.  However, by adding a
  suitable multiple of the identity we can always make the Hamiltonian
  traceless, and so there is no loss of generality in making this
  assumption.} \emph{control Hamiltonian} $H(t)$.  It is convenient to
expand the control Hamiltonian in terms of the generalized Pauli
matrices, which we take to be the set of $n$-fold tensor products of
the single-qubit Pauli matrices, omitting $I^{\otimes n}$.  The
resulting expansion is:
\begin{eqnarray} \label{eq:control}
  H(t) = \sum_\sigma \gamma^\sigma(t) \, \sigma.
\end{eqnarray}
Note that we always omit the term $\sigma = I^{\otimes n}$ from such
sums.  The functions $\gamma^\sigma(t)$ are known as \emph{control
  functions}; the $(4^n-1)$-dimensional vector $\gamma(t)$ whose
entries are the individual control functions $\gamma^\sigma(t)$ is
known as \emph{the} control function.  Sometimes it is convenient to
omit the $t$ and just write $\gamma$ to denote the entire
vector-valued control function $\gamma(t)$.

In order that $U$ be correctly synthesized, Schr\"odinger's equation
requires that the control function $\gamma(t)$ satisfies the
equations:
\begin{eqnarray} \label{eq:control-problem}
  \frac{dV}{dt} = -i H(t) V; \,\,\,\, V(0) = I; \,\,\,\, V(1) = U.
\end{eqnarray}
We have chosen $t=0$ as the initial time, and $t = 1$ as the time at
which we desire the evolution to reach $U$.  These choices are
arbitrary, and it is not difficult to prove that the definition given
below of the cost of synthesizing $U$ does not depend on the values
chosen for these times.

%
%
It is helpful to assume that $\gamma(t)$ is a smooth (i.e.,
$C^\infty$) function of $t$. We say that a smooth control function
$\gamma$ satisfying Equations~(\ref{eq:control})
and~(\ref{eq:control-problem}) is a \emph{valid} control function
generating $U$.

\subsection{Cost functions and right-invariant local metrics on manifolds}
\label{subsec:cost-function-and-metrics}

Our goal is to determine the most efficient way of generating $U$.  To
make the notion of efficiency precise, we introduce the \emph{cost}
$c_f(\gamma)$ associated to a valid control function $\gamma$,
\begin{eqnarray}
  c_f(\gamma) \equiv \int_0^1 dt \, f(\gamma(t)),
\end{eqnarray}
where $f : \mathbb R^{4^n-1} \rightarrow \mathbb R$ is a real-valued
function of the control function $\gamma(t)$.  We study below the
properties which $f$ ought to have if $c_f(\gamma)$ is to be a good
measure of efficiency.

We now define the \emph{cost} $c_f(U)$ of the unitary $U$ as the
infimum of the cost $c_f(\gamma)$ over all valid control functions
$\gamma$ generating $U$,
\begin{eqnarray}
  c_f(U) \equiv \inf_\gamma c_f(\gamma).
\end{eqnarray}
Note that we will refer to all three of $f, c_f(\gamma)$ and $c_f(U)$
as the cost function, depending on context.

The remainder of this subsection is devoted to arguing that the cost
function $f$ is equivalent to a geometric object known as a
right-invariant local metric on the manifold $SU(2^n)$.  

To make this argument, in~\ref{subsubsec:desired} we begin by noting a
few properties that $f$ ought to have, if $c_f(\gamma)$ and $c_f(U)$
are to be good measures of cost.  The purpose here is simply to
\emph{motivate} the list of properties we will demand of $f$, and so
the discussion focuses on heuristic arguments and intuition building,
rather than on rigorous proofs.

Our list of desired properties for $f$ in hand,
in~\ref{subsubsec:reformulation} we move to the framework of
differential geometry, and show that with these properties, $f$
corresponds to a right-invariant local metric on $SU(2^n)$.  The
advantage of moving to this geometric viewpoint is that it allows the
well-developed tools and viewpoint of geometry to be applied.

\subsubsection{Desired properties of the cost function $f$}
\label{subsubsec:desired}

\textbf{Continuity:} A background assumption useful in our later
arguments is that $f$ be continuous.  Obviously this is reasonable on
physical grounds.

\textbf{Positivity:} Given the interpretation of $c_f(U)$ as the cost
of synthesizing $U$, we expect that $c_f(U) \geq 0$, with equality if
and only if $U = I$, the identity operation.  Using the continuity of
$f$, it is straightforward to see that this is equivalent to the
condition $f(y) \geq 0$, with equality if and only if $y = 0$.

\textbf{Positive homogeneity:} Physically, if we double the intensity
of the Hamiltonian for a while, but halve the time it is applied, we
wouldn't expect the cost to change, as the total effort required is
the same.  Mathematically, this idea may be expressed by the
requirement that $f$ be positively homogeneous, i.e., that $f( \alpha
y) = \alpha f(y)$, for any positive real number $\alpha$, and any
vector $y$.

\textbf{Achievement of the infimum:} Another useful background
assumption is that the infimum in the definition of $c_f(U)$ is
achieved by some valid control function $\gamma$.  This is not
strictly necessary for the arguments we make below, but it does
streamline them.  

\textbf{The triangle inequality:} We will argue that $f$ ought to
satisfy the triangle inequality, $f(x+y) \leq f(x)+f(y)$.  Suppose
$\gamma$ is the control function which minimizes $c_f(U)$.  Fix $t$,
and suppose there exist $x$ and $y$ such that $\gamma(t) = x+y$ and
$f(x+y) > f(x) + f(y)$.  Choose a value $\Delta > 0$ sufficiently
small that $f(\gamma(s))$ is effectively constant over the interval $s
\in [t,t+\Delta]$.  We construct a new modified control function
$\gamma_1$ which takes the value $2 x$ on the interval $[t,t+\Delta /
2]$, the value $2 y$ on the interval $[t+\Delta/2, t+\Delta]$, and
otherwise takes the same values as $\gamma$.  Note that this function
is not valid, since it is neither smooth nor does it exactly satisfy
Equations~(\ref{eq:control}) and~(\ref{eq:control-problem}).  However,
it is easy to regularize $\gamma_1$ to produce a control function
$\gamma_2$ that is valid, and has essentially the same cost.  It
follows that $c_f(\gamma_2) = c_f(\gamma_1) < c_f(\gamma)$, which
contradicts the presumed minimality of $\gamma$.  This suggests that
$f$ should satisfy the triangle inequality $f(x+y) \leq f(x)+f(y)$ for
all $x$ and $y$.  

It is worth noting the additional point that if we desire our minimal
curves to be unique then the triangle inequality needs to be strict,
i.e., $f(x+y) \leq f(x)+f(y)$, with equality if and only if $x$ and
$y$ are along the same ray emanating from the origin.  If the strict
triangle inequality is not satisfied, then $\gamma$ could be modified
to produce a different valid control function $\gamma_2$ with the same
cost, using an argument similar to that above.  This observation will
be useful later, in our discussion of Finsler geometry.

Summarizing, we have argued that the cost function ought to satisfy
the conditions: $f(y) \geq 0$ with equality iff $y=0$; $f$ is
positively homogeneous, i.e., $f(\alpha y) = \alpha f(y)$ for all
positive $\alpha$; and the triangle inequality $f(x+y) \leq
f(x)+f(y)$.  We now show that these conditions imply a correspondence
between $f$ and right-invariant local metrics on $SU(2^n)$.

\subsubsection{Geometric reformulation of the cost function}
\label{subsubsec:reformulation}

We assume the reader is familiar with the concepts of elementary
differential geometry, and merely review the necessary notation and
nomenclature.  The reader unfamiliar with any of these concepts is
advised to consult an introductory text such as Isham~\cite{Isham99a}
or Lee~\cite{Lee03a}.  It will also help to have some familiarity with
Riemannian geometry --- also covered, albeit rather more briefly, in
those texts --- but this is not as essential.

We denote a smooth (i.e., $C^\infty$) $n$-dimensional manifold by $M$;
we typically omit ``smooth'' and just refer to $M$ as a manifold.  We
will often denote points on $M$ by $x$, and local co-ordinate systems
by $\phi: S \rightarrow \mathbb R^n$, where $S$ is an open subset of
$M$, and $\phi$ is a homeomorphism of $S$ into a subset of $\mathbb
R^n$.  The tangent space to $M$ at point $x$ is denoted $T_xM$, and we
often use $y$ to denote a vector in a tangent space such as $T_xM$.
We write $(x,y)$ to denote an element of the tangent bundle $TM$.  We
will have much interest in $C^\infty$ maps $f: M \rightarrow N$
between manifolds $M$ and $N$, where by $C^\infty$ we mean that all
the derivatives of the map $f$ exist and are continuous with respect
to any local co-ordinate systems on $M$ and $N$.  We will use the term
$C^\infty$ interchangeably with the term ``smooth''.  A curve on $M$
is a smooth map $s : I \rightarrow M$, where $I$ is an interval in
$\mathbb{R}$.  Given such a smooth map $f : M \rightarrow N$, and
fixing $x \in M$, we use $f_* : T_x M \rightarrow T_{f(x)} N$ to
denote the natural pushforward map connecting the tangent spaces $T_x
M$ and $T_{f(x)} N$.

We define a \emph{manifold with local metric} as a manifold $M$
equipped with a function $F : TM \rightarrow [0,\infty)$ such that for
each fixed $x$, the function $F(x,y)$ satisfies: $F(x,y) \geq 0$ with
equality iff $y = 0$; $F(x,y)$ is positively homogeneous in $y$; and
$F(x,y)$ satisfies the triangle inequality in the second variable.
$F$ is called the local metric.

So far as I am aware the term ``local metric'' is not a standard term
in geometry, however in this paper we'll find it a useful unifying
term that can be specialized to give the standard concepts of a
Finsler or Riemannian metric.

We define the \emph{length} of a curve $s : I \rightarrow M$ on a
manifold $M$ with local metric $F$ by
\begin{eqnarray} \label{eq:length-definition}
  l_F(s) \equiv \int_I dt \, F(s(t),[s]_t),
\end{eqnarray}
where $[s]_t \in T_{s(t)} M$ is the tangent to $s$ at $s(t)$.  With
this definition the length is invariant under reparameterization of
the curve. More precisely, suppose $\phi : I \rightarrow I'$ is a
smooth, strictly monotone increasing function taking the interval $I$
onto another interval $I'$.  Then the identity $l_F(s) = l_F(s \circ
\phi)$ follows easily from the definition of length, the positive
homogeneity of the local metric, i.e., $F(x,\alpha y) = \alpha
F(x,y)$, elementary differential geometry, and calculus.

We define the \emph{distance} $d_F(x,x')$ between two points $x$ and
$x'$ on $M$ as the infimum of $l_F(s)$ over all curves $s$ connecting
$x$ and $x'$.

To explain the connection between local metrics and the cost function,
we need to introduce \emph{locally adapted co-ordinates} on the
manifold $SU(2^n)$.  We define these co-ordinates as follows.  First,
fix an \emph{origin} $U \in SU(2^n)$, and define $\psi : \mathbb
R^{4^n-1} \rightarrow SU(2^n)$ by $\psi(x) \equiv \exp(-i x \cdot
\sigma) U$, where $\sigma$ here is the $(4^n-1)$-component vector
whose entries are the generalized Pauli matrices.  This maps $\mathbb
R^{4^n-1}$ onto $SU(2^n)$ in a many-to-one fashion.  Supposing $S$ is
an open subset of $\mathbb R^{4^n-1}$ such that $\psi : S \rightarrow
SU(2^n)$ is one-to-one, we define a set of \emph{$U$-local adapted
  co-ordinates} to be the inverse function $\phi : \psi(S) \rightarrow
S$.

In practice, we shall only be interested in $U$-local adapted
co-ordinates for unitary matrices $V$ in some small neighbourhood of
$U$.  In such a neighbourhood we may simply define
\begin{eqnarray}
  \phi(V)^\sigma \equiv \frac{i \, \mbox{tr}(\ln(VU^\dagger) \sigma)}{2^n},
\end{eqnarray}
where $\ln$ is the standard branch of the logarithm.  We call this
co-ordinate system \emph{the} $U$-local adapted co-ordinates.  Note
that we will use the terms ``$U$-local adapted co-ordinates'' and
``local adapted co-ordinates'' interchangeably, with the former
preferred when we wish to be specific about the identity of the
origin, and the latter preferred when we wish to omit specific
identification of the origin.  

Any co-ordinate system $x^\sigma$ for an open neighbourhood of $U \in
SU(2^n)$ induces a corresponding \emph{natural} co-ordinate system for
the tangent space $T_U SU(2^n)$.  This is done by singling out the
\emph{natural basis} $(\partial / \partial x^\sigma)_U$ for $T_U
SU(2^n)$, and expanding an arbitrary tangent vector $y \in T_U
SU(2^n)$ as $y = \sum_\sigma y^\sigma (\partial / \partial
x^\sigma)_U$.  We refer to the $y^\sigma$ as the \emph{natural
  co-ordinates} for $y$ with respect to the co-ordinate system
$x^\sigma$.  We refer to the co-ordinate system $(x^\sigma,y^\sigma)$
for $T \, SU(2^n)$ as a natural co-ordinate system for the tangent
bundle $T \, SU(2^n)$.

Suppose now that $V$ is a unitary for which $U$-local adapted
co-ordinates are defined. Then those co-ordinates give rise to a set
of natural co-ordinates for $T_V SU(2^n)$, which we call the
\emph{natural $U$-adapted co-ordinates}, or just the \emph{natural
  adapted co-ordinates}, when it is clear what value $U$ takes.  The
following proposition gives a way of computing the natural adapted
co-ordinates for the vector tangent to a curve.

\begin{proposition} {} \label{prop:tangent-adapted}
  Let $U(t)$ be a smooth curve in $SU(2^n)$.  Then (a) $i
  \frac{dU}{dt} U^\dagger$ is Hermitian, and (b) the natural
  $U(t)$-adapted co-ordinates $y^\sigma$ for the tangent to the curve,
  $[U]_t \in T_{U(t)} SU(2^n)$, are determined by the
  equation\footnote{Note the convention, used here and throughout,
    that the $\sigma$ in $y^\sigma$ refers to a specific generalized
    Pauli matrix, while in $y \cdot \sigma$ it refers to the entire
    vector of generalized Paulis.}
  \begin{eqnarray} \label{eq:tangent-adapted}
    y \cdot \sigma = i \frac{dU}{dt} U^\dagger.
  \end{eqnarray}
  In particular, we have $y^\sigma = i \mbox{tr}(\sigma \, dU/dt \,
  U^\dagger) / 2^n$.
\end{proposition}

\textbf{Proof:} There is a sense in which (a) follows from (b), since
the $y^\sigma$ are real, by definition.  Nonetheless, it seems
worthwhile to include the following brief proof of (a).  Using the
unitarity of $U(t)$ we have
\begin{eqnarray}
  0 = \frac{d(U U^\dagger)}{dt} = \frac{dU}{dt}U^\dagger + 
  U \frac{dU}{dt}^\dagger.
\end{eqnarray}
Thus $(dU/dt) U^\dagger$ is anti-Hermitian, which proves part~(a).

To prove (b), we expand
\begin{eqnarray}
  U(t+\Delta) & = & U(t) + \Delta \left. \frac{dU}{dt}\right|_t + 
  O(\Delta^2) \\
  & = & \exp\left( -i\times i \left. \frac{dU}{dt}\right|_t 
    U(t)^\dagger \Delta \right) U(t) \nonumber \\
  & & + O(\Delta^2). 
\end{eqnarray}
It follows that the natural $U(t)$-adapted co-ordinates of the tangent
$[U]_t$ are determined by Equation~(\ref{eq:tangent-adapted}).
\qed

Suppose now that we define a local metric $F$ on the manifold
$SU(2^n)$ by $F(U,y) \equiv f(\gamma)$, where $f$ is a cost function,
and $\gamma$ is the vector whose co-ordinates are the natural
$U$-adapted co-ordinates of $y$.  With this definition, the conditions
for $F$ to be a local metric follow immediately from the conditions we
obtained earlier for the cost function $f$ --- positivity, positive
homogeneity, and the triangle inequality.

Furthermore, observe from Proposition~\ref{prop:tangent-adapted} that
if $V(t)$ is a solution to Equations~(\ref{eq:control})
and~(\ref{eq:control-problem}), then the natural $V(t)$-adapted
co-ordinates for the tangent to the curve $V$ at the point $V(t)$ are
just the control functions $\gamma^\sigma(t)$.  It follows that the
cost $c_f(\gamma)$ is equal to the length $l_F(V)$ of the curve $V$ on
the manifold $SU(2^n)$, and therefore that the cost $c_f(U)$ is equal
to the distance $d_F(I,U)$ between the identity operation $I$ and the
unitary $U$.

We can also show that $F$ is an example of a special type of local
metric known as a right-invariant local metric.  In general, suppose
$G$ is any Lie group (such as, for example, $SU(2^n)$), and $F : TG
\rightarrow [0,\infty)$ is a local metric defined on $G$.  Then $F$ is
\emph{right-invariant} if $F(x,r_{x*}(y)) = F(e, y)$, where $e$ is the
Lie group identity, $r_x$ is right multiplication by $x$, i.e.,
$r_x(x') = x' x$, and $r_{x*}$ is the pushforward of $r_x$ at $e$,
mapping $T_e G$ to $T_x G$.  With this definition, we see that
right-invariant local metrics are simply those which are constant in
natural adapted co-ordinates, and thus our $F$ is an example of a
right-invariant local metric.  We note for later use that
left-invariant local metrics are defined similarly to right-invariant
local metrics, except the invariance is now under left multiplication,
and that a local metric which is both left- and right-invariant is
said to be \emph{bi-invariant}.

Summarizing, we have argued on general grounds that the cost
associated to the synthesis of a unitary operation $U$ ought to be
given by the distance $d_F(I,U)$ between the identity $I$ and $U$, for
some right-invariant local metric $F$.

\subsection{Minimal curves and lower bounds}
\label{subsec:minimal_curves}

In this subsection we prove a simple theorem relating geometry to
minimal size quantum circuits, giving sufficient conditions for a
local metric $F$ to satisfy $d_F(I,U) \leq m_{\cal G}(U)$.

To state the theorem, it helps to first introduce a little more
notation and nomenclature.  Suppose ${\cal G}$ is some set of unitary
gates which is exactly universal when acting on $n$ qubits.  For
example, ${\cal G}$ might consist of all one- and two-qubit unitary
gates which can be written in the form $\exp(-i \alpha \sigma)$, where
$0 \leq \alpha \leq 1$ and $\sigma$ is either a single-qubit Pauli, or
a two-qubit Pauli.

Suppose ${\cal H}$ is a set of Hermitian matrices such that the map
${\cal H} \rightarrow {\cal G}$ defined by $H \rightarrow \exp(-i H)$
is one-to-one and onto.  As an example corresponding to the set ${\cal
  G}$ defined in the previous paragraph, we have ${\cal H}$ consisting
of all Hermitian matrices of the form $\alpha \sigma$, where $\alpha$
and $\sigma$ are as in the definition of ${\cal G}$.

On the tangent bundle $T SU(2^n)$ we write $(V,H)$ to denote the pair
consisting of $V \in SU(2^n)$ and the tangent vector $[\exp(-i H t)
V]_{t=0} \in T_V SU(2^n)$.  This notation should not be confused with
the similar notation $(V,y)$ for elements of $T SU(2^n)$ that we've
used up to now, and will continue to use when appropriate, where $V
\in SU(2^n)$ and $y \in T_V SU(2^n)$.  The advantage of the new
notation is that it allows us to write $F(V,H)$ to denote the cost of
applying the Hamiltonian $H$ at the point $V \in SU(2^n)$.

Finally, suppose $F$ is a local metric on $SU(2^n)$ satisfying $F(V,H)
\leq 1$ for all $V$ in $SU(2^n)$, and for all $H$ in ${\cal H}$.  Then
we say that $F$ is \emph{${\cal G}$-bounding}.  The reason for this
nomenclature is provided by the following theorem, which shows that
whenever $F$ is ${\cal G}$-bounding, $d_F(I,U)$ provides a lower bound
on $m_{\cal G}(U)$.

\begin{theorem} {} \label{thm:lower-bounds}
  Suppose ${\cal G}$ is an exactly universal gate set on $SU(2^n)$,
  and ${\cal H}$ is a corresponding set of Hermitian matrices, as
  described above.  Suppose $F$ is a ${\cal G}$-bounding local metric
  on $SU(2^n)$.  Then for any fixed $U$ in $SU(2^n)$ the inequality
  \begin{eqnarray}
    d_F(I,U) \leq m_{\cal G}(U)
  \end{eqnarray}
  holds.
\end{theorem}

\textbf{Proof:} Suppose that $U_1 = \exp(-i H_1), \ldots, U_{m_{\cal
    G}(U)} = \exp(-i H_{m_{\cal G}(U)})$ is a minimal sequence of
quantum gates synthesizing $U$, where the gates are chosen from ${\cal
  G}$.  We define a curve $V(t)$ between $I$ and $U$ by defining a
control function induced by this gate sequence.  The definition of the
control function is:
\begin{eqnarray}
  \frac{\gamma(t) \cdot \sigma}{m_{\cal G}(U)} \equiv
  \left\{ \begin{array}{ll}
  H_1 & \mbox{ if } 0 \leq t < 1/m_{\cal G}(U) \\
  H_2 & \mbox{ if } 1/m_{\cal G}(U) \leq t < 2/m_{\cal G}(U) \\
  \ldots & \ldots \\
  H_{m_{\cal G}(U)} & \mbox{ if } 
  1-1/m_{\cal G}(U) \leq t \leq 1.
\end{array}
\right.
\end{eqnarray}
This control function gives rise to a curve between $I$ and $U$ by
integrating Equations~(\ref{eq:control})
and~(\ref{eq:control-problem}).  However, the curve is not smooth, and
so the control function is not valid.  To correct this we regularize
$\gamma$ to produce a smooth control function $\gamma_1$ that also
generates $U$.

We do this regularization using a real-valued smooth function $r(t)$
with the properties that: (a) $r(t) = 0$ for any point $t$ which is an
integer multiple of $1/m_{\cal G}(U)$; (b) $r(t) \geq 0$; and (c) for
any integer $j$ the integral of $r(t)$ over the interval $[j/m_{\cal
  G}(U),(j+1)/m_{\cal G}(U)]$ is $1/m_{\cal G}(U)$.  Such a function
is easily constructed using the standard techniques of analysis.

We now define a modified control function $\gamma_1(t) \equiv r(t)
\gamma(t)$, and the corresponding curve $V(t)$ is defined by
integrating Equations~(\ref{eq:control})
and~(\ref{eq:control-problem}).  This is now a smooth curve connecting
$I$ and $U$, and the length of the curve is:
\begin{eqnarray}
  l_F(V) & = & \int_0^1 dt \, F(V(t),\gamma_1(t) \cdot \sigma) \\
  & = & \int_0^1 dt \, r(t) F(V(t), \gamma(t) \cdot \sigma) \\
  & \leq & \int_0^1 dt \, r(t) m_{\cal G}(U) \\
  & = & m_{\cal G}(U),
\end{eqnarray}
where we have applied, respectively: the definition of length; the
property that a local metric is positively homogeneous in the second
variable; the fact that $\gamma(t) \cdot \sigma / m_{\cal G}(U)$ is in
${\cal H}$, and the assumption $F(V,H) \leq 1$ for all $V \in SU(2^n)$
and $H \in {\cal H}$; and, finally, the fact that for any integer $j$
the integral of $r(t)$ over the interval $[j/m_{\cal
  G}(U),(j+1)/m_{\cal G}(U)]$ is $1/m_{\cal G}(U)$.  It follows that
$d_F(I,U) \leq m_{\cal G}(U)$, as claimed.  \qed

\subsection{Geodesics and Finsler geometry}
\label{subsec:finsler-definition}

We have argued that the cost function $f$ corresponds to a
right-invariant ${\cal G}$-bounding local metric $F$ on $SU(2^n)$.  In
this subsection we will argue that if we are to study the function
$d_F(I,U)$ using the calculus of variations, then $F$ ought to belong
to a special class of local metrics known as Finsler metrics.

To see this, we again start with some heuristic motivating arguments
regarding the properties of $f$, before turning to a discussion of
what these properties mean geometrically, i.e., in terms of the local
metric $F$.

\textbf{Smoothness:} In order to apply the calculus of variations, we
need to make some smoothness assumptions about the cost function $f$.
Although it is not strictly necessary, we will assume that the cost
function is differentiable to all orders away from the origin, i.e.,
$f(y)$ is a $C^\infty$ function, except at the origin.  The reason we
exclude the origin from the smoothness requirement is that if $f$ is
non-negative and positively homogeneous, as we argued it ought to be
earlier, then the only way $f$ can be differentiable at the origin is
if it vanishes everywhere.

\textbf{The strict triangle inequality and the Hessian:} We argued
earlier that $f$ ought to satisfy the strict triangle inequality.  Our
assumption that $f$ is also smooth enables us to recast the strict
triangle inequality in a more convenient and (almost) equivalent form.
We define the $(4^n-1) \times (4^n-1)$ \emph{Hessian} matrix whose
entries are $H_{\sigma \tau} \equiv \frac{1}{2} \partial^2 f^2 /
\partial y^\sigma \partial y^\tau$, where $y^\sigma$ is our notation
for the $\sigma$th co-ordinate slot in the function $f^2$.  It turns
out that a necessary condition for the triangle inequality to hold is
that the Hessian matrix be a positive matrix.  A sufficient condition
for the strict triangle inequality to hold is that the Hessian matrix
be strictly positive, and this is the condition we shall impose on
$f$.

These conditions are well-understood in the Finsler geometry
community, and so we merely outline why these facts are the case,
omitting the details.  The interested reader is referred to, e.g.,
Section~1.2 of the book by Bao, Chern and Shen~\cite{Bao00a} for a
more detailed discussion.  Our reason for including this brief
discussion here is partially motivational, but also because many of
the ideas introduced will be needed later, when we discuss the
approximation of local metrics by Finsler metrics.

The key concept behind these results is that of the \emph{indicatrix}.
The indicatrix of $f$, denoted $S_f$, is defined to consist of all
those points $y$ such that $f(y) = 1$.  The indicatrix generalizes the
unit sphere, where $f$ is the Euclidean norm function.  We define the
\emph{unit ball} $B_f$ for $f$ to consist of all those points $y$ such
that $f(y) \leq 1$.

Assuming $f$ is positively homogeneous, it is not difficult to show
that the triangle inequality $f(x+y) \leq f(x)+f(y)$ is equivalent to
the condition that the unit ball $B_f$ be convex.  Under the same
condition, the strict triangle inequality is easily seen to be
equivalent to the condition that $B_f$ be \emph{strictly} convex,
i.e., any line joining two points of $B_f$ should be contained
entirely within the interior of $B_f$, except possibly at the
endpoints.  Equivalently, the tangent hyperplane to $B_f$ at any point
of $S_f$ should only touch a single point of $B_f$.

How does this geometry relate to the Hessian?  Suppose we pick a point
$y_0$ on the indicatrix.  Consider the tangent plane, defined to
consist of those points $y$ satisfying $\nabla f \cdot y = \nabla f
\cdot y_0$.  Define $\Delta \equiv y-y_0$.  Expanding $f(y) =
f(y_0+\Delta)$ in a Taylor series in $\Delta$ and doing some
elementary manipulations gives
\begin{eqnarray}
  f(y_0+\Delta) = 1 + \Delta^T H \Delta + O(\Delta^3),
\end{eqnarray}
where $\Delta^T$ indicates the transpose of $\Delta$, and $H$ is the
Hessian matrix.  Thus, provided the Hessian is strictly positive, it
follows that $y_0$ is the only point in the tangent hyperplane which
is also in $B_f$, and so the the indicatrix is strictly convex, and
the strict triangle inequality is satisfied.

The standard terminology is that $f$ is \emph{strongly convex} when
the Hessian is strictly positive; this is a stronger condition than
strict convexity of $f$.  It is not difficult to find examples where
the Hessian is only positive, not positive definite, and yet the
strict triangle inequality holds. Essentially, at such points the
quadratic terms in $f(y_0+\Delta)$ may vanish, yet we still have
$f(y_0+\Delta) > 1$, due to the contribution of higher-order terms,
ensuring the strict triangle inequality holds.  See, e.g.,
Exercise~1.2.7 of~\cite{Bao00a}.

We will see that there are significant advantages to assuming that the
Hessian is strictly positive, i.e., that strong convexity holds.  In
particular, in Section~\ref{sec:geodesic_equation} we'll see that this
is exactly the condition needed to make the geodesic equation a second
order differential equation. If the indicatrix is strictly but not
strongly convex then the geodesic equation is not a second order
differential equation, but one must instead go to higher order
equations, which substantially complicates the study of geodesics.

Summarizing, we have argued that the cost function $f$ ought to be
smooth away from the origin, and the Hessian of $f$ ought to be
strictly positive definite.  We now show that this means that the
corresponding local metric $F$ on $SU(2^n)$ is a Finsler metric.

Finsler geometry is a well-developed subject, and our treatment here
is based on the standard text by Bao, Chern and Shen~\cite{Bao00a},
and on the notes of \'Alvarez and Dur\'an~\cite{Alvarez98a}, to which
the reader should refer for more details.

To define Finsler metrics it helps to first define the notion of a
Minkowski norm, which is a function $N : \mathbb{R}^d \rightarrow
[0,\infty)$ which is smooth away from the origin, satisfies $N(y) \geq
0$ with equality if and only if $y = 0$, is positively homogeneous,
and strongly convex in the same sense described earlier, i.e., the
Hessian matrix $H = (H_{jk})$ whose components are the partial
derivatives
\begin{eqnarray} H_{jk} \equiv
\frac{1}{2} \frac{\partial N^2}{\partial y^j \partial y^k}
\end{eqnarray}
is strictly positive when evaluated at any point $y \in \mathbb{R}^d$.

Informally, a Finsler metric is a family of Minkowski norms on the
tangent spaces to the manifold, one norm for each point on the
manifold, and such that the Minkowski norms vary smoothly as a
function of position on the manifold.  More precisely, a \emph{Finsler
  metric} on a manifold $M$ is a function $F : TM \rightarrow
[0,\infty)$ such that $F(x,y)$ is a smooth function of $x$ and $y$ for
all $x$ and all $y \neq 0$, and such that for each fixed $x$,
$F(x,\cdot)$ is a Minkowski norm on the tangent space $T_x M$.

Clearly, Finsler metrics are a special case of local metrics.  Note
also that Riemannian metrics are a special case of Finsler metrics,
coinciding with the condition that $F(x,\cdot)^2$ be a quadratic form
for each fixed $x \in M$.

When the cost function $f$ is smooth away from the origin and has a
strictly positive Hessian, we see that the corresponding local
metric\footnote{Recall that the components of $\gamma$ are just the
  natural $U$-adapted co-ordinates for $y$.}  $F(U,y) \equiv
f(\gamma)$ is an example of a right-invariant Finsler metric.  Thus,
the local metrics we shall be most interested in studying in this
paper are right-invariant ${\cal G}$-bounding Finsler metrics.

Finsler metrics have a number of useful properties that we note here
without proof.

First, the Hopf-Rinow theorem (see page~168 and exercise~6.2.11 on
page~155 of~\cite{Bao00a}) implies that for a compact Finsler manifold
the infimum in the definition of $d_F$ is always achieved by some
smooth curve.

Second, Euler's theorem for smooth and homogeneous functions (see
Section~1.2 of~\cite{Bao00a}) implies a number of useful identities
satisfied by Finsler metrics:
\begin{eqnarray} \label{eq:homogeneity-1}
  \sum_j \frac{\partial F^2}{\partial y^j} y^j & = & 2 F^2 \\
  \label{eq:homogeneity-2}
  \sum_{jk} \frac{\partial^2 F^2}{\partial y^j \partial y^k} y^j y^k & = & 
  2F^2
  \\ \label{eq:homogeneity-3}
  \sum_j \frac{\partial F^2}{\partial y^j \partial y^k \partial y^l} 
  y^j & = & 0.
\end{eqnarray}
In these equations, the $y^j$ are any fixed set of co-ordinates for
the tangent space $T_x M$.  Note that
Equation~(\ref{eq:homogeneity-3}) can be recast in several different
ways, depending on which order the partial derivatives are taken.  We
use several of these different orderings later.

\subsection{Examples of local metrics whose minimal length curves provide
  lower bounds on circuit size}
\label{subsec:examples}

We have argued that if $F$ is to be useful for proving lower bounds on
$m_{\cal G}(U)$ then it ought to be a ${\cal G}$-bounding local
metric; even better, a Finsler metric, in order that the calculus of
variations and results like the Hopf-Rinow theorem be applicable.  In
addition to these properties, a local metric $F$ ideally should have
the following three properties: (1) it is easy to determine the
minimal curve length $d_F(I,U)$; (2) there exist families of unitaries
indexed by $n$ and with long minimal geodesics according to $F$, i.e.,
families of unitaries for which $d_F(I,U)$ scales exponentially with
$n$; and (3) $d_F(I,U)$ is polynomially equivalent to $m_{\cal G}(U)$.
Note that~(3) implies~(2), since unitaries for which $m_{\cal G}(U)$
is exponential are known to exist~\cite{Knill95a}.

It is a significant open problem to find a local metric satisfying all
of these properties.

The purpose of the present subsection is to introduce four natural
candidates for such a local metric, denoted $F_1, F_2, F_p$, and
$F_q$, and to discuss the extent to which they satisfy these desired
properties.  All of them are right-invariant ${\cal G}$-bounding local
metrics, and thus at least satisfy the inequality $d_F(I,U) \leq
m_{\cal G}(U)$.

We will see that one of these local metrics, $F_2$, definitely does
not have all the desired properties.  Although our discussion is not
conclusive, it is plausible that each of the other three local metrics
does possess the desired properties, with the caveat that $F_1$ and
$F_p$ are not Finsler, and must be approximated by suitable Finsler
metrics.  In particular, we will present some heuristic evidence that
$F_1$ and $F_p$ satisfy all our criteria.

To define our local metrics let $U \in SU(2^n)$, $y \in T_U SU(2^n)$,
and suppose $y$ has natural $U$-adapted co-ordinates $y^\sigma$, and
so can be thought of as corresponding to the Hamiltonian $y \cdot
\sigma$.  Then we define
\begin{eqnarray}
  F_1(U,y) & \equiv & \sum_\sigma |y^\sigma| \\
  F_2(U,y) & \equiv & \sqrt{ \sum_\sigma (y^\sigma)^2} \\
  F_p(U,y) & \equiv & \sum_\sigma p(\mbox{wt}(\sigma)) |y^\sigma| \\
  F_q(U,y) & \equiv & \sqrt{ \sum_\sigma q(\mbox{wt}(\sigma)) (y^\sigma)^2 }.
\end{eqnarray}
In these expressions, $\mbox{wt}(\sigma)$ is the Hamming weight of the
Pauli matrix $\sigma$, and $p(\cdot)$ and $q(\cdot)$ are \emph{penalty
  functions} that penalize the control function whenever Pauli terms
of high weight contribute to the control Hamiltonian.  E.g., we might
choose $p(j) = 4^j$ to provide an exponential penalty for the use of
higher-weight Pauli matrices.  We return to the choice of the penalty
function below.

As was remarked earlier, it is often useful to write $F(U,H) \equiv
F(U,y)$, where $H$ is the Hamiltonian such that $y = [\exp(-i Ht)
U]_{t=0}$.  With this convention it is easily verified that
right-invariant local metrics such as $F_1, F_2, F_p$ and $F_q$ have
no $U$-dependence, and so we sometimes write $F(H) \equiv F(\cdot,H)$.
Note that $F(H)$ is a norm on $su(2^n)$.

We now apply Theorem~\ref{thm:lower-bounds} to these example metrics.
To do this, we choose the universal gate set ${\cal G}$ as in the
example described earlier, specifically, to consist of all one- and
two-qubit unitary gates which can be written in the form $\exp(-i
\alpha \sigma)$, where $0 \leq \alpha \leq 1$ and $\sigma$ is a Pauli
matrix of weight one or two.  The corresponding ${\cal H}$ consists of
all Hermitian matrices of the form $\alpha \sigma$.

We see immediately that with these choices $F_1$ and $F_2$ satisfy the
hypothesis of Theorem~\ref{thm:lower-bounds}, namely, we have
$F_1(V,H) \leq 1$ and $F_2(V,H) \leq 1$ for all $V \in SU(2^n)$ and
all $H \in {\cal H}$.  Thus, we have $d_{F_1}(I,U) \leq m_{\cal G}(U)$
and $d_{F_2}(I,U) \leq m_{\cal G}(U)$.  In the case of $F_p$ and $F_q$
we need the supplementary assumptions that $p(1), p(2) \leq 1$ and
$q(1), q(2) \leq 1$, respectively.  With these assumptions it is
easily verified that the hypothesis of Theorem~\ref{thm:lower-bounds}
holds, and so we have $d_{F_p}(I,U) \leq m_{\cal G}(U)$ and
$d_{F_q}(I,U) \leq m_{\cal G}(U)$.

\subsubsection{Properties of $F_2$ and $F_q$}

$F_2$ is an example of a very well understood class of local metrics:
it is a bi-invariant Riemannian metric on $SU(2^n)$.  Such local
metrics have the nice property that their geodesics are completely
understood --- they are the curves of the form $\exp(-i Ht)$, where $H
\in su(2^n)$ --- as are properties such as curvature and other
geometric invariants.  See, e.g., the end-of-chapter problems in
Chapter~5 of~\cite{Lee97a}.

Although $F_2$ is well understood, it turns out that the lower bounds
on $m_{\cal G}(U)$ obtained from $F_2$ are at best constant, and thus
are not especially interesting. In particular, we will show that for
any $U$ we have $d_{F_2}(I,U) \leq \pi^2$, and thus the best possible
bound we can hope for is $\pi^2 \leq m_{\cal G}(U)$.

To see that $d_{F_2}(I,U) \leq \pi^2$, observe first the identity
$F_2(V,H) = \mbox{tr}(H^2)/2^n$.  Then for any $U$, select Hermitian
$H$ with eigenvalues in the range $-\pi$ to $\pi$ and such that
$\exp(-iH) = U$.  Define a curve $\gamma$ from $I$ to $U$ via
$\gamma(t) \equiv \exp(-i H t)$; this can actually be shown to be the
minimal length geodesic through $U$, although we won't need this fact.
The length of this curve is $l_{F_2}(\gamma) = \int_0^1 dt \,
F_2(\exp(-iHt),H) = \mbox{tr}(H^2)/2^n$.  It follows that\footnote{As
  this is the minimal length curve, the inequality which follows is,
  in fact, an equality.}  $d_{F_2}(I,U) \leq
\frac{\mbox{tr}(H^2)}{2^n}$.  But from the eigenvalue bounds on $H$ it
follows that $\mbox{tr}(H^2) \leq \pi^2 2^n$, whence we obtain
\begin{eqnarray}
  d_{F_2}(I,U) \leq \pi^2.
\end{eqnarray}
Thus we can never hope to prove any more than that $\pi^2 \leq m_{\cal
  G}(U)$ using $F_2$.  

The essential reason $F_2$ is unsuitable for proving lower bounds is
that it contains no information about the tensor product structure, as
can be seen from the expression $F_2(H) = \mbox{tr}(H^2)/2^n$.  How
can we encode information about the tensor product structure in the
metric, in order to have some hope of obtaining non-constant lower
bounds on circuit size?  One possibility is to simply exclude the
possibility of the control Hamiltonian containing Pauli terms of
weight higher than two.  To do this we need to move to the field of
sub-Riemannian geometry~\cite{Montgomery02a}, which is concerned with
the situation where there are restrictions on the allowed directions
that a curve may take in the tangent space.  This direction of
research is under investigation by the author.

Another possible approach is to introduce a penalty function
$q(\cdot)$ which penalizes the use of high weight Pauli matrices in
the control Hamiltonian.  Many forms for the penalty function suggest
themselves, and it is not clear which, if any, is the most
appropriate.  Here is one illustrative choice:
\begin{eqnarray} \label{eq:example-penalty}
  q(j) = \left\{ \begin{array}{ll} 1 & \mbox{ if } j = 1 \mbox{ or } 2\\
      k & \mbox{otherwise},
    \end{array}
    \right.
\end{eqnarray}
where $k$ is a penalty that may depend on the number of qubits $n$,
but is otherwise constant.  As $k$ becomes large we expect that this
approach will yield essentially the same geodesics as in the
sub-Riemannian approach mentioned above.  One advantage of using such
a local metric is that it is a right-invariant Riemannian metric, and
such local metrics are quite well understood.  See, e.g., Appendix~2
of~\cite{Arnold89a}, and the end-of-chapter problems in Chapter~5
of~\cite{Lee97a}.

\subsubsection{Properties of $F_1$ and $F_p$}

The local metric $F_1$ is perhaps the most promising cost function,
due to the following interpretation.  Suppose $U$ may be generated by
applying sequentially the Pauli matrices $\sigma_1, \sigma_2, \ldots$
for times $t_1,t_2,\ldots$.  Then the length $l_{F_1}$ of the
corresponding curve is just the total time $t_1+t+2+ \ldots$.  An
approximate converse is also true.  In particular, using the Trotter
formula it is easy to prove that for any Hamiltonian $H$ and time $t
\geq 0$ it is possible to approximate $\exp(-i Ht)$ arbitrarily well
using just Pauli Hamiltonians $\sigma$, applied for a total time
$F_1(H)t$.  It follows that given a curve $\gamma$, we can approximate
$\gamma$ arbitrarily well using a sequence of evolutions, each one a
Hamiltonian evolution with Hamiltonian some generalized Pauli sigma
matrix, with the total time of evolution being just $l_{F_1}(\gamma)$.
Note also that Hamiltonian evolution according to any single
generalized Pauli matrix is easily simulatable with at most a linear
number of gates in the standard quantum circuits model.

Thus, $d_{F_1}(I,U)$ has a natural interpretation as the minimal time
required to generate $U$ by switching between Hamiltonians chosen from
the set of generalized Pauli matrices, each of which can be
efficiently simulated in the standard quantum circuits model.

It is tempting to suppose on the basis of this interpretation that
$d_{F_1}(I,U)$ must be polynomially equivalent to $m_{\cal G}(U)$.
Although I believe this likely to be the case, it is possible to
imagine, for example, that $d_{F_1}(I,U)$ is small (maybe even a
constant), and yet the oscillations in any near-optimal path $\gamma$
are so wild that to approximate it in the quantum circuit model
requires exponentially many gates.  If this were the case then the
cost in doing the computation would not be due to the actual
Hamiltonian evolution, but rather due to extremely frequent switching
between very short evolutions by different Pauli Hamiltonians.

In the event that this turns out to be the case, one potential
resolution would be to work instead with $F_p$, in which a penalty
function $p(\cdot)$ is used to penalize the use of higher-weight Pauli
matrices in the Hamiltonian evolution.  As was the case for $F_q$ it
is not clear what the best choice of penalty function is, but various
simple alternatives naturally suggest themselves.  In particular, as
for $F_q$, by choosing $p$ appropriately we can effectively rule out
some directions of movement on the manifold.  When $F_p$ is by
approximated by a suitable Finsler metric (as described in the next
paragraph) ruling out such directions of movement places us
effectively in the realm of ``sub-Finsler'' geometry~\cite{Lopez00a,
  Clelland04a}.

The main disadvantage of $F_1$ and $F_p$ is that they are not Finsler
metrics, and thus we can't directly apply the calculus of variations
to study their minimal length curves.  To remedy this situation, in
Appendix~\ref{app:approximating-metrics} we explain how to approximate
$F_1$ and $F_p$ by a family of Finsler metrics $F_{1\Delta}$ and
$F_{p\Delta}$, where $\Delta > 0$ is a small parameter such that as
$\Delta \rightarrow 0$, $F_{1\Delta} \rightarrow F_1$ and $F_{p\Delta}
\rightarrow F_p$.  More precisely, we show that
\begin{eqnarray}
  F_1(U, y) & \leq & F_{1\Delta}(U, y) \leq \frac{F_1(U, y)}{1-(4^n-1)\Delta},
  \\
  F_p(U, y) & \leq & F_{p\Delta}(U, y) \leq \frac{F_p(U, y)}{1-P\Delta},
\end{eqnarray}
where $P \equiv \sum_\sigma p(\mbox{wt}(\sigma))$.  Thus, provided
$\Delta$ is sufficiently small the Finsler metrics $F_{1\Delta}$ and
$F_{p\Delta}$ provide excellent approximations to $F_1$ and $F_p$,
respectively.  As a result, our strategy for understanding the minimal
length curves of $F_1$ and $F_p$ is to study them via the geodesics of
$F_{1\Delta}$ and $F_{p\Delta}$.

\subsubsection{Summary and comparison}

We have introduced four classes of local metric, $F_1,F_2,F_p$, and
$F_q$.  All four provide lower bounds on $m_{\cal G}(U)$ through the
inequality $d_F(I,U) \leq m_{\cal G}(U)$.  Summarizing and comparing
their various properties:
\begin{enumerate}
\item $F_2$ is capable of producing at best a constant lower bound on
  $m_{\cal G}(U)$, and thus is not especially interesting.
  
\item $F_q$ is a modified version of $F_2$ in which we introduce a
  penalty for the application of higher-weight Pauli Hamiltonians.
  The main advantages of $F_q$ are that it is easy to compute and
  Riemannia.  It is also straightforward to compute quantities such as
  curvature and the Christoffel symbols using standard results about
  right-invariant Riemannian metrics.
 
\item $F_1$ is the best motivated of all the four local metrics.  In
  particular, we showed that $d_{F_1}(I,U)$ is the minimal time
  required to synthesize $U$ using some set of efficiently simulatable
  Hamiltonians.  It can be approximated arbitrarily well using a
  suitable Finsler metric $F_{1\Delta}$.
  
\item $F_p$ is a modified version of $F_1$ in which we introduce a
  penalty for the application of higher-weight Pauli Hamiltonians.  It
  can also be approximated arbitrarily well using a suitable Finsler
  metric $F_{p\Delta}$.

\end{enumerate}
I conjecture that for suitable choices of the penalty functions, $p$
and $q$, all three of the local metrics $F_1, F_p$, and $F_q$ are
polynomially equivalent to $m_{\cal G}(U)$, and thus could, in
principle, be used to prove exponential lower bounds on $m_{\cal
  G}(U)$.  In this paper we will not resolve the correctness of this
conjecture, although Section~\ref{subsec:CVP} presents some evidence
that $F_1$ (and thus $F_p$) has exponential length minimal curves,
which provides some indirect evidence for the conjecture.

Despite the lack of a proof of this conjecture, it remains an
interesting problem to understand the geodesic structure of each of
these classes of local metric, and what this implies about the
distances $d_F(I,U)$.  It is to this problem that we turn for the
remainder of this paper.

\section{Computing the geodesic equation}
\label{sec:geodesic_equation}

In this section we'll explain how to explicitly construct the geodesic
equation for each of the Finsler metrics we have introduced.  This
equation is a second-order differential equation whose solutions are
geodesics of the Finsler manifolds, i.e., curves in $SU(2^n)$ which
are local extrema of the Finsler length.  

The exact form of the geodesic equation is rather complex, even for
the simplest of our local metrics; we shall not write it out
explicitly.  Our goal in this section is to describe a general
procedure which can be used to compute the geodesic equation, and thus
enable numerical and analytic investigation of geodesics.  A detailed
numerical investigation of the geodesic structure is underway and will
appear elsewhere.

We begin in Subsection~\ref{subsec:review_geodesic_equation}, with a
brief review of the geodesic equation for Finsler geometry.  This is
standard material in Finsler geometry, and so we cover it quickly,
merely outlining derivations, and referring the reader to standard
references such as~\cite{Bao00a} for more details.

In order to apply the geodesic equation it is most convenient to pick
out a single co-ordinate system\footnote{In fact,no single co-ordinate
  system can cover all of $SU(2^n)$.  However, it is possible to pick
  a co-ordinate system that covers all of $T SU(2^n)$ except a set of
  measure zero, and that is what we will do.  We return later to the
  question of what to do on the set of measure zero.} for $SU(2^n)$,
and carry out all calculations with respect to those co-ordinates.  In
particular, using a single set of co-ordinates greatly facilitates
integration of the geodesic equation, and analytic investigation of
that equation.  Unfortunately, the Finsler metrics we have introduced
in Subsection~\ref{subsec:examples} are all defined in terms of local
adapted co-ordinates, which vary from point to point on $SU(2^n)$.

To remedy this situation, the majority of this section is taken up
with learning how to change from local adapted co-ordinates to a
single fixed set of co-ordinates for $T SU(2^n)$, which we call
natural Pauli co-ordinates.  We explain how to make this change of
variables for the special case of $T SU(2)$ in
Subsection~\ref{subsec:su2}.  In Subsection~\ref{subsec:su2n} we
describe the main ideas behind the change of variables for $T
SU(2^n)$, before describing some convenient calculational techniques
for making this change in
Subsection~\ref{subsec:explicit-computation-su2n}.

With these results in hand it is possible in principle to explicitly
compute the geodesic equation for each of the local metrics we have
introduced.  In practice, the actual form of the equation is rather
complicated, and it is more convenient to investigate solutions
numerically, or using techniques such as the analysis of symmetries.
Indeed, as we do not do any numerical analysis in this paper, later
sections of the paper in fact depend only on the basic form of the
geodesic equation, presented in
Subsection~\ref{subsec:review_geodesic_equation}, and so the other
parts of this section may be skipped if the reader's main interest is
in the geodesic solutions constructed in
Section~\ref{sec:geodesic_properties}.

The section also contains a brief digression, in
Subsection~\ref{subsec:direct_sum}, whose purpose is to illustrate the
results of Subsection~\ref{subsec:review_geodesic_equation} with some
simple results on the effects ancilla qubits have on minimal length
curves.  In particular, we show that for a suitable family of Finsler
metrics, there is a neighbourhood of the identity in which for all
unitaries $U$ the distance $d_F(I,U)$ is equal to the distance $d_F(I
\otimes I,U \otimes I)$, i.e., the distance is unaffected by the
addition of ancilla on which the unitary acts trivially.  These
results, simple as they are, represent our only progress on the
problem of understanding the effect ancilla qubits have on minimal
size quantum circuits.

\subsection{General form of the geodesic equation}
\label{subsec:review_geodesic_equation}

In this subsection we construct the geodesic equation.  We follow the
standard procedure used in Finsler geometry (see, e.g.~\cite{Bao00a})
to construct the equation, and for that reason we merely outline the
relevant calculations.  

In order to construct the geodesic equation, it is convenient to fix a
set of co-ordinates for $SU(2^n)$.  We will label these co-ordinates
$x^j$, and the corresponding natural co-ordinates for the tangent
space $y^j$.  The co-ordinate system $x^j$ chosen is a completely
arbitrary chart from among the atlas of possible co-ordinates on the
manifold.  Note that at any point along a curve $s = s(t)$ the tangent
vector has natural co-ordinates given by $y^j = dx^j / dt$.

In general, the co-ordinates $x^j$ do not cover all of $SU(2^n)$.  As
a result, to construct geodesics it is in general necessary to change
the co-ordinate system being used as the geodesic moves across the
manifold.  However, for this initial discussion it is most convenient
to imagine that the co-ordinate system has been fixed, and we are
computing geodesics that lie within the region covered by that
co-ordinate system.  

%
%
Recall that we defined the length of a curve $s : I \rightarrow M$ by
$l_F(s) \equiv \int_I dt \, F(s(t),[s]_t)$, where $[s]_t \in
T_{s(t)}M$ is the vector tangent to $s$ at $t$.  In terms of the
co-ordinates $x^j$ and $y^j = dx^j / dt$ this may be rewritten $l_F(s)
= \int_I dt F(x,y)$, where $F = F(x,y)$ is $F$ expressed in terms of
the co-ordinates $x = (x^j)$ and the corresponding natural
co-ordinates $y = (y^j) = (dx^j/dt)$ for the tangent vector $[s]_t$.

In order to determine the geodesics which minimize the length we use
the calculus of variations, a review of which may be found
in~\cite{Gelfand63a}.  It is a standard result in the calculus of
variations that any curve $s$ which is an extremum of the functional
$\int_I dt \, F(x,y)$ must satisfy the Euler-Lagrange equations:
\begin{eqnarray} \label{eq:Euler-Lagrange}
  \frac{d}{dt} \left( \frac{\partial F}{\partial y^j} \right) =
  \frac{\partial F}{\partial x^j}.
\end{eqnarray}
We will sometimes refer to this equation as the geodesic equation, for
its solutions give rise to geodesics.  However, for practical purposes
it is more convenient to recast the geodesic equation in other forms.
First, and rather remarkably, it is possible to replace $F$ by $F^2$
in Equation~(\ref{eq:Euler-Lagrange}) to get an equation which is
essentially equivalent:
\begin{eqnarray} \label{eq:Euler-Lagrange-2}
  \frac{d}{dt} \left( \frac{\partial F^2}{\partial y^j} \right) =
  \frac{\partial F^2}{\partial x^j}.
\end{eqnarray}
To understand this equivalence and its significance, note the
easily-verified identity:
\begin{eqnarray} 
  & & \left( \frac{d}{dt} \left( \frac{\partial F^2}{\partial y^j} \right) -
  \frac{\partial F^2}{\partial x^j} \right) \nonumber \\
  & = & 
  2 \frac{dF}{dt} \frac{\partial F}{\partial y^j}
  + 2 F \left( \frac{d}{dt} \left( \frac{\partial F}{\partial y^j} \right) -
  \frac{\partial F}{\partial x^j} \right).  \label{eq:EL-equivalence}
\end{eqnarray}
Suppose $s = s(t)$ is a curve which solves the geodesic equation,
Equation~(\ref{eq:Euler-Lagrange}).  Then we can reparameterize this curve
to give an equal length curve $\tilde s$ which has constant speed,
i.e., $dF/dt = 0$ along the curve $\tilde s$.  It is straightforward
to verify that the curve $\tilde s$ also solves the geodesic equation,
Equation~(\ref{eq:Euler-Lagrange}).  However, since $dF/dt = 0$, we see
from Equation~(\ref{eq:EL-equivalence}) that $\tilde s$ is also a solution
to Equation~(\ref{eq:Euler-Lagrange-2}).

Conversely, suppose the curve $s = s(t)$ is a solution to
Equation~(\ref{eq:Euler-Lagrange-2}).  Then we have:
\begin{eqnarray}
  \frac{dF^2}{dt} & = & \sum_j \left( \frac{\partial F^2}{\partial x^j} 
    \frac{dx^j}{dt} + \frac{\partial F^2}{\partial y^j} 
    \frac{dy^j}{dt} \right) \\
  & = & \sum_j \frac{d}{dt} \left( \frac{\partial F^2}{\partial y^j} y^j 
  \right).
\end{eqnarray}
Using Equation~(\ref{eq:homogeneity-1}) we see that $\sum_j
\frac{\partial F^2}{\partial y^j} y^j = 2F^2$, and so the previous
equation implies $dF^2 / dt = 2 \, dF^2 / dt$, and so any solution to
Equation~(\ref{eq:Euler-Lagrange-2}) automatically satisfies the
constant speed condition $dF^2 / dt = 0$.

Summing up, the class of curves which solve
Equation~(\ref{eq:Euler-Lagrange}) is equivalent to the class of
curves which solve Equation~(\ref{eq:Euler-Lagrange-2}), up to a
reparameterization which leaves the length invariant, and thus is of
no interest.  However, solutions to
Equation~(\ref{eq:Euler-Lagrange-2}) have the additional useful
property that they are automatically curves of constant speed.  We
therefore refer to either Equation~(\ref{eq:Euler-Lagrange}) or
Equation~(\ref{eq:Euler-Lagrange-2}) as the geodesic equation,
depending on context.

Equation~(\ref{eq:Euler-Lagrange-2}) may be recast in an equivalent
form analogous to the standard geodesic equation for Riemannian
manifolds, Equation~(\ref{eq:geodesic}).  First, using
Equations~(\ref{eq:homogeneity-2}) we substitute $F^2 = g_{lm} y^l
y^m$, where we define $g_{lm} \equiv \frac{1}{2} \frac{\partial
  F^2}{\partial y^l \partial y^m}$ to be the Hessian, and we use the
summation convention that repeated indices are summed over, unless
otherwise stated.  The right-hand side of
Equation~(\ref{eq:Euler-Lagrange-2}) then becomes $g_{lm,x^j} y^l
y^m$, where we use the subscript notation ``$_{,x^j}$'' to indicate a
partial derivative with respect to the $x^j$ co-ordinate.

To analyse the left-hand side of Equation~(\ref{eq:Euler-Lagrange-2}), we
again substitute $F^2 = g_{lm} y^l y^m$, and apply the usual rules of
calculus together with Equation~(\ref{eq:homogeneity-3}) to obtain
\begin{eqnarray}
  \frac{\partial F^2}{\partial y^j} = 2g_{jm} y^m.
\end{eqnarray}
Taking the total derivative with respect to $t$, and again using
Equation~(\ref{eq:homogeneity-3}), we obtain
\begin{eqnarray}
  \frac{d}{dt} \left( \frac{\partial F^2}{\partial y^j} \right)
  = 2g_{jm,x^n} y^m y^n + 2 g_{jm} \frac{d y_m}{dt}.
\end{eqnarray}
Using these results we see that the geodesic equation,
Equation~(\ref{eq:Euler-Lagrange-2}), may be recast in the form
\begin{eqnarray} \label{eq:EL2-recast}
  2g_{jm,x^n} y^m y^n + 2 g_{jm} \frac{d y_m}{dt} = g_{lm,x^j} y^l y^m.
\end{eqnarray}
By assumption, the matrix whose components are the $g_{jk}$ is
strictly positive, and so it is possible to define a matrix $g^{jk}$
which is the inverse of $g_{jk}$.  Multiplying
Equation~(\ref{eq:EL2-recast}) by this inverse and doing some
rearrangement, we obtain our alternate form of the geodesic equation,
\begin{eqnarray} \label{eq:alternate-form-geodesic-equations}
  \frac{d^2 x^j}{dt^2} + \Gamma^j_{kl} \frac{dx^k}{dt}
\frac{dx^l}{dt} = 0,
\end{eqnarray}
where the Christoffel symbols $\Gamma^j_{kl}$ are defined by
\begin{eqnarray}
  \Gamma^j_{kl} \equiv \frac{g^{jm}}{2} \left( g_{mk,x^l}
    + g_{ml,x^k}-g_{kl,x^m} \right).
\end{eqnarray}
Formally, this definition for the Christoffel symbols appears
identical to that used in Riemannian geometry.  The difference is that
in Riemannian geometry the $g_{jk}$ are functions of $x$ alone, while
in Finsler geometry they are typically functions of $y$ as well.

Summing up, we have presented the geodesic equation in three different
(but equivalent) forms,
Eqs.~(\ref{eq:Euler-Lagrange}),~(\ref{eq:Euler-Lagrange-2}), and
(\ref{eq:alternate-form-geodesic-equations}).  The latter is
explicitly in the form of a second order differential equation, and so
the usual existence and uniqueness theorems for second order ordinary
differential equations apply.  In particular, given an initial
position and velocity (i.e., tangent vector) on the manifold, the
remainder of the geodesic is completely specified by
Equation~(\ref{eq:alternate-form-geodesic-equations}).  Of course,
this form of initial data problem is not the problem of most interest
to us.  We are more concerned with the problem of studying geodesics
where two points on the geodesic are specified, but the initial
velocity is unknown.

\subsection{Application: Ancilla qubits and direct sum theorems}
\label{subsec:direct_sum}

As an illustration of the results of the previous subsection, consider
the problem of determining $m_{\cal G}(U \otimes V)$, where $U$ is a
unitary on an $n_A$-qubit system, labeled $A$, and $V$ is a unitary on
an $n_B$-qubit system, labeled $B$.  An interesting question is to ask
how $m_{\cal G}(U \otimes V)$ is related to $m_{\cal G}(U)$ and
$m_{\cal G}(V)$, where the notation ${\cal G}$ is overloaded in the
obvious way.  It is clear that $m_{\cal G}(U \otimes V) \leq m_{\cal
  G}(U) + m_{\cal G}(V)$.  Might this inequality sometimes be strict,
or must it be satisfied with equality?

Questions like this are the province of \emph{direct sum theorems},
which seem to have first been considered by~\cite{Yao79a}.
Essentially, a direct sum theorem seeks to establish whether a set of
two or more computational tasks can be collectively accomplished using
fewer resources than the sum of the resources required for the
individual tasks.

Of particular interest in the context of the current paper is the case
$V = I$, which is related to the problem of determining whether or not
ancilla can help in implementing $U$.

With the tools available we cannot directly study the behaviour of
$m_{\cal G}(U \otimes V)$, but we can study the related question of
whether $d_{F_{AB}}(I_A \otimes I_B,U \otimes V) = d_{F_A}(I_A,U) +
d_{F_B}(I_B,V)$, where $F_A, F_B$ and $F_{AB}$ are suitable Finsler
metrics on the respective spaces.  We will not solve this problem in
general, but can easily obtain some simple results indicating that, at
least near the identity, this equality will always be satisfied for
suitable choices of the Finsler metrics.

Suppose $F_A,F_B$ and $F_{AB}$ are Finsler metrics on $SU(2^{n_A}),
SU(2^{n_B})$ and $SU(2^{n_A+n_B})$, respectively.  We say they form an
\emph{additive triple} of Finsler metrics if:
\begin{eqnarray}
  F^2_{AB}(U\otimes V, H_A+H_B) = F^2_A(U,H_A)+F^2(V,H_B), \nonumber \\
\end{eqnarray}
where $H_A \in su(2^{n_A})$, $H_B \in su(2^{n_B})$, and we abuse
notation by omitting tensor factors which act trivially, like $I_A
\otimes \cdot $ and $\cdot \otimes I_B$.

Suppose $U(t)$ is a geodesic of $F_A$, and $V(t)$ is a geodesic of
$F_B$.  If $F_A, F_B$ and $F_{AB}$ are an additive triple of Finsler
metrics, then it follows from the linearity of the geodesic equation,
Equation~(\ref{eq:Euler-Lagrange-2}), that $W(t) \equiv U(t) \otimes
V(t)$ is a geodesic of $F_{AB}$.

An example of an additive triple of Finsler metrics is the triple
$F_{q_A}, F_{q_B}$, and $F_{q_{AB}}$, where the penalty functions
$q_A, q_B$ and $q_{AB}$ satisfy the consistency condition $q_A(j) =
q_B(j) = q_{AB}(j)$, for all $j$ where this condition is well-defined.
It follows that if $U(t)$ and $V(t)$ are geodesics of $F_{q_A}$ and
$F_{q_B}$, then $U(t) \otimes V(t)$ is a geodesic of $F_{q_{AB}}$.

We have seen that a tensor product of geodesic curves is itself a
geodesic curve, for additive triples of Finsler metrics.  Can we
conclude that the shortest curve connecting $I_A \otimes I_B$ and $U
\otimes V$ is just the tensor product of the shortest curve connecting
$I_A$ and $U$ with the shortest curve connecting $I_B$ and $V$?

I do not know if this is the case, in general.  However, a well-known
theorem of Finsler geometry (see, e.g., Section~6.3 of~\cite{Bao00a})
asserts that for any manifold $M$ equipped with a Finsler metric $F$,
there exists a constant $r > 0$ such that any geodesic of length less
than $r$ is guaranteed to be a minimal length curve.

It follows that for $U$ and $V$ in some finite-size neighbourhood of
the respective identity operations, $I_A$ and $I_B$, the minimal curve
from $I_A \otimes I_B$ to $U \otimes V$ is just the tensor product of
the minimal curves in the respective spaces.  It follows that in this
neighbourhood the distance $d_{F_{AB}}(I_A \otimes I_B,U \otimes V)$
is equal to the sum $d_{F_A}(I_A,U)+d_{F_B}(I_B,V)$.

Specializing to the case where $V = I_B$, we see that for all $U$ in
some finite-size neighbourhood of $I_A$ the minimal length curve from
$I_A \otimes I_B$ to $U \otimes I_B$ is guaranteed to be exactly equal
to the minimal length curve from $I_A$ to $U$.  That is, there is a
neighbourhood of the identity in which the presence of ancilla does
not help in shortening the length of the minimal curves.

\subsection{Pauli co-ordinates}

As noted in the introduction to this section, our primary goal in this
section is to explain how to compute the geodesic equation with
respect to a fixed co-ordinate system on $SU(2^n)$.  The co-ordinates
we shall use are the Pauli co-ordinates.  In our earlier language,
Pauli co-ordinates are $I$-adapted local co-ordinates for $SU(2^n)$,
where $I$ is the $n$-qubit identity operation.  The unitary
corresponding to Pauli co-ordinates $x$ is given by
\begin{eqnarray}
  \exp\left(-i x \cdot \sigma \right) = \exp\left(-i \sum_\sigma 
  x^\sigma \sigma \right).
\end{eqnarray}
Inverting, the co-ordinates $x^\sigma$ corresponding to some unitary
$U$ are given by
\begin{eqnarray}
  x^\sigma \equiv \frac{i \, \mbox{tr}(\ln(U) \sigma)}{2^n},
\end{eqnarray}
where $\ln$ is some branch of the logarithm.  We will be particularly
interested in the case where $\ln$ is the standard branch of the
logarithm, defined around a branch cut along the negative real axis.
We call these co-ordinates the \emph{standard Pauli co-ordinates}, or
just Pauli co-ordinates.  Note that the standard Pauli co-ordinates
are defined for any unitary operator whose spectrum does not include
$-1$, and thus are defined everywhere in $SU(2^n)$ except on a set of
measure zero.

Just as for local adapted co-ordinates, the Pauli co-ordinates on
$SU(2^n)$ give rise to natural co-ordinates on the tangent space $T_U
SU(2^n)$.  In the remainder of this section we will typically use
$x^\sigma$ to denote Pauli co-ordinates, $\tilde x^\sigma$ to denote
local adapted co-ordinates, and $y^\sigma$ and $\tilde y^\sigma$ to
denote the corresponding natural co-ordinates on $T_U SU(2^n)$.

\subsection{Changing co-ordinates in $T_U SU(2)$}
\label{subsec:su2}

Let's begin with the example of $T_U SU(2)$, where it is relatively
straightforward to change between the natural Pauli and natural
locally adapted co-ordinates.  The key to making the change is the
following theorem.  Note that in this subsection (and in the
associated Appendix~\ref{app:qubit-change-vars}) we will find it
useful to work both with vectors in $\mathbb R^3$, and with vectors
relating directly to objects in the tangent bundle $T_U SU(2)$.  We
will refer to the former using the notations $\vec x, \vec y$ and
$\tilde y$, while for the latter we will use $x$ to refer to the
vector of Pauli co-ordinates for $U \in SU(2)$, and $y$ to refer to an
element of $T_U SU(2)$.

\begin{theorem} {} \label{thm:change-qubit-coords}
  Fix $\vec x \in \mathbb R^3$.  Then
  \begin{eqnarray} \label{eq:change-qubit-coords}
    & & \exp(-i (\vec x + t \vec y) \cdot \sigma) \nonumber \\
   & = &
    \exp(-i t \tilde y \cdot \sigma) \exp(-i \vec x \cdot 
    \sigma)+O(t^2),
  \end{eqnarray}
  where $\vec y$ may be expressed as a function of $\vec x$ and
  $\tilde y$:
  \begin{eqnarray} \label{eq:y-qubit}
    \vec y = \tilde y_{\|} + \| \vec x \| \cot(\| \vec x \|) 
    \tilde y_\perp + \tilde y \times  \vec x.
  \end{eqnarray}
  In this expression $\hat x \equiv \vec x / \| \vec x\|$ is the
  normalized vector in the $\vec x$ direction, $\tilde y_{\|} \equiv
  \hat x \cdot \tilde y \, \hat x$ is the component of $\tilde y$ in
  the $\hat x$ direction, and $\tilde y_\perp \equiv \tilde y - \tilde
  y_{\|}$ is the component of $\tilde y$ orthogonal to the $\hat x$
  direction.  We can invert this equation to express $\tilde y$ in
  terms of $\vec x$ and $\vec y$, obtaining:
  \begin{eqnarray} \label{eq:tilde-y-qubit}
    \tilde y =  \vec y_{\|} + \sinc(2\|\vec x\|)  \vec y_\perp 
    +\sinc^2(\| \vec x\|)  \vec x
    \times  \vec y_\perp,
  \end{eqnarray}
  where $\vec y_{\|}$ is now the component of $\vec y$ in the $\hat x$
  direction, $\vec y_\perp$ is the component of $\vec y$ orthogonal to
  $\hat x$, and $\sinc(z) \equiv \sin(z)/z$ is the standard sinc
  function.
\end{theorem}

The proof of this theorem is a straightforward calculation.  We
describe the details in Appendix~\ref{app:qubit-change-vars}.
Alternately, it is a useful and not entirely trivial exercise to
deduce the theorem from the more general results about $SU(2^n)$ in
the next subsection.

To see how this theorem enables us to change co-ordinates, fix $U \in
SU(2)$, and fix $y \in T_U SU(2)$.  Suppose $x$ are the Pauli
co-ordinates for $U$.  Then we have
\begin{eqnarray}
  y = \sum_\sigma y^\sigma \left( \frac{\partial}{\partial x^\sigma}
    \right)_U
\end{eqnarray}
for some set of co-ordinates $y^\sigma$, and where $( \partial /
\partial x^\sigma)_U$ are the natural Pauli co-ordinate basis vectors
for $T_U SU(2)$.  Setting $\vec x \equiv x$ and letting $\vec y$ have
components $y^\sigma$, we see that the vector $y \in T_U SU(2)$ is
tangent to the curve $\exp(-i (\vec x + t\vec y) \cdot
\sigma)$ at $t=0$.  That is:
\begin{eqnarray}
  y = [ \exp(-i(\vec x+t \vec y) \cdot \sigma) ]_{t=0}.
\end{eqnarray}
Applying Theorem~\ref{thm:change-qubit-coords}, we obtain
\begin{eqnarray}
  y = [ \exp(-i t \tilde y \cdot \sigma) U+O(t^2) ]_{t=0}.
\end{eqnarray}
Neglecting the terms of order $t^2$ does not change the tangent at $t
= 0$, and so
\begin{eqnarray}
  y = [ \exp(-i t \tilde y \cdot \sigma) U ]_{t=0}.
\end{eqnarray}
It follows that $\tilde y^\sigma$ are the natural adapted co-ordinates
for $T_U SU(2)$.  Thus Theorem~\ref{thm:change-qubit-coords} relates
the natural Pauli co-ordinates $y^\sigma$ on $T_U SU(2)$ to the
natural $U$-adapted co-ordinates $\tilde y^\sigma$ on $T_U SU(2)$.

\subsection{Changing co-ordinates in $T_U SU(2^n)$}
\label{subsec:su2n}

The key result enabling the change between natural Pauli and natural
locally adapted co-ordinates is a generalization of
Theorem~\ref{thm:change-qubit-coords} which applies to unitary
operators in arbitrary dimensions.  This result is due to
Baker~\cite{Baker1905a}, Campbell~\cite{Campbell1897a,Campbell1897b}
and Hausdorff~\cite{Hausdorff06a}, and we refer to it as the \emph{BCH
  formula}.  Note that this result is \emph{not} what is usually
referred to as the BCH formula by physicists, but is a generalization.
See Section~3.4 of~\cite{Reutenauer93a} for a recent discussion of the
BCH formula and related results.
  
To state the BCH formula it helps to first define linear
superoperators (i.e., linear operations on matrices) $\ad_X$ and
${\cal I}$ by $\ad_X(Z) \equiv [X,Z]$, and ${\cal I}(Z) \equiv Z$.
With these definitions we have the following.

\begin{theorem}[BCH formula] {} \label{thm:general-change-vars}
  Suppose $X$ and $Y$ are $d \times d$ Hermitian operators.  Then we
  have
  \begin{eqnarray}
    \exp(-i t \tilde Y) \exp(-i X) = \exp(-i(X+tY)) + O(t^2),
  \end{eqnarray}
  where the $d \times d$ Hermitian operator $\tilde Y$ is defined by
  \begin{eqnarray}
    \tilde Y \equiv {\cal E}_X(Y),
  \end{eqnarray}
  and ${\cal E}_X$ is a linear superoperator defined by
  \begin{eqnarray}
    {\cal E}_X & = & \frac{\exp(-i \ad_X)-{\cal I}}{-i \ad_X}.
  \end{eqnarray}
\end{theorem}

In this theorem, the superoperator ${\cal E}_X$ is understood as a
formal power series.  In particular, the operator $\ad_X$ is not
invertible, so strictly speaking the expression given for ${\cal E}_X$
is not even defined.  Nonetheless, treating the expressions as power
series, we see that
\begin{eqnarray} \label{eq:power-series}
  {\cal E}_X & = & \sum_{j=0}^{\infty} \frac{(-i \ad_X)^j}{(j+1)!}
\end{eqnarray}
is defined and everywhere convergent.  We will discuss in the next
subsection how to explicitly calculate the action of ${\cal E}_X$ in a
convenient fashion.  For now we take it as given that this can be
done, and discuss how Theorem~\ref{thm:general-change-vars} allows us
to change variables in $T_U SU(2^n)$.

The discussion, of course, follows similar lines to the discussion in
the previous subsection.  We fix $y \in T_U SU(2^n)$, and suppose
$\vec x$ is a vector of Pauli co-ordinates for $U$, so $U = \exp(-i
\vec x \cdot \sigma)$.  We will find it convenient to use $x^\sigma$
both to denote Pauli co-ordinates, and also the particular Pauli
co-ordinates for $U$, with the meaning to be determined by context.
Then we have
\begin{eqnarray}
  y & = & \sum_\sigma y^\sigma \left( \frac{\partial}
    {\partial x^\sigma} \right)_U \\
  & = & \left[ \exp(-i(\vec x + t \vec y)\cdot \sigma \right]_{t=0},
\end{eqnarray}
where $\vec y$ is the vector whose entries are the natural Pauli
co-ordinates $y^\sigma$.  Using the BCH formula we have
\begin{eqnarray}
y  & = & \left[ \exp(-it \tilde y \cdot \sigma) U + O(t^2) \right]_{t=0} \\
  & = & \left[ \exp(-it \tilde y \cdot \sigma) U \right]_{t=0},
\end{eqnarray}
where $\tilde y$ is determined by $\vec x$ and $\vec y$ through the
equations
\begin{eqnarray} 
  \label{eq:general-change-coords-pt1}
  \tilde y \cdot \sigma & = & {\cal E}_{\vec x \cdot \sigma}
  ( \vec y \cdot \sigma) \\
  \label{eq:general-change-coords-pt2}
  {\cal E}_{\vec x \cdot \sigma} & = & 
  \frac{\exp(-i \ad_{\vec x \cdot \sigma})-{\cal I}}
    {-i \ad_{\vec x \cdot \sigma}}.
\end{eqnarray}
Note that the components of $\tilde y$ may be extracted from this
expression by multiplying both sides by some specific Pauli matrix
$\sigma$ and taking the trace.

Equations~(\ref{eq:general-change-coords-pt1})
and~(\ref{eq:general-change-coords-pt2}) are general equations telling
us how to transform from natural Pauli co-ordinates in $T_U SU(2^n)$
to natural adapted co-ordinates in $T_U SU(2^n)$.  By applying the
inverse operation we can transform from natural adapted co-ordinates
into natural Pauli co-ordinates.  A straightforward but somewhat
lengthy calculation shows that in the case where $n = 1$ these results
reduce to the results for $SU(2)$ deduced in the previous section.  We
omit the details of this calculation.

\subsection{Explicit computation of the change of co-ordinates}
\label{subsec:explicit-computation-su2n}

In the last subsection we explained how to change from natural Pauli
co-ordinates to natural adapted co-ordinates in $T_U SU(2^n)$, through
Equations~(\ref{eq:general-change-coords-pt1})
and~(\ref{eq:general-change-coords-pt2}).  These formulas are compact,
but it is not entirely evident how to perform an explicit calculation
of this change of co-ordinates.  In this subsection we explain how to
perform such calculations, and also how to do the inverse change, from
natural adapted co-ordinates to natural Pauli co-ordinates.  This
enables, in principle, the explicit computation of all terms in the
geodesic equation,
Equation~(\ref{eq:alternate-form-geodesic-equations}).

One way of performing such calculations is to expand ${\cal E}_{\vec x
  \cdot \sigma}$ in a power series in $\ad_{\vec x \cdot \sigma}$, as
specified by Equation~(\ref{eq:power-series}).  Computations can then
be carried out to a good approximation simply by computing the first
few terms of the power series.  Along similar lines, the inverse to
${\cal E}_{\vec x \cdot \sigma}$ also has a power series expansion
(see the discussion in Section~3.4 of~\cite{Reutenauer93a}), and so
computations of the inverse can be carried out along similar lines.

However, there is a useful alternate approach offering the possibility
of exact computation, which we now describe.  In particular, we will
describe a general method to compute ${\cal E}_X(Y)$.

There are two main difficulties facing us in the computation.  The
first is that it is computationally inconvenient to deal with
superoperators like $\ad_X$.  To alleviate this difficulty we will
\emph{vectorize} our expressions.  Vectorization is a procedure that
converts operators into vectors, and superoperators into operators, to
obtain equivalent expressions involving only vectors and ordinary
operators.  This step is not strictly necessary, but it is extremely
convenient.  As the vectorization formalism is not entirely standard
in the quantum information literature, an introduction to this
formalism is presented in Appendix~\ref{app:vectorizing}, to which the
reader not familiar with vectorization should now turn.

The second, and more serious, difficulty, is that $\ad_X$ is not
invertible.  Our solution is to decompose $Y$ into a component lying
in the kernel of $\ad_X$, and a component lying in the orthocomplement
of the kernel.  We then compute the action of ${\cal E}_X$ on each of
these two spaces separately.  This computation is also greatly
facilitated by use of the vectorization formalism.

We begin with a useful characterization of the kernel of $\ad_X$,
which we denote $\mbox{ker}(\ad_X)$.  Recall that $\mbox{ker}(\ad_X)$
consists of all those matrices $Z$ such that $\ad_X(Z) = 0$.  The
following proposition gives a computationally convenient description
for $\mbox{ker}(\ad_X)$.

\begin{proposition}
  Suppose the Hermitian matrix $X$ has spectral decomposition $\sum_j
  x_j P_j$, where the $x_j$ are the distinct eigenvalues of $X$, and
  the $P_j$ project onto the corresponding eigenspaces.  Then the
  operation
  \begin{eqnarray}
    {\cal P}(Z) \equiv \sum_j P_j Z P_j
  \end{eqnarray}
  projects onto the kernel of $\ad_X$.  
\end{proposition}

\textbf{Proof:} To see that ${\cal P}$ is a projector we need to show
that it is Hermitian and satisfies ${\cal P}\circ {\cal P} = {\cal
  P}$.  Both of these facts are easily verified.
  
To complete the proof we need to show that ${\cal P}(Z) = Z$ if and
only if $Z$ is an element of the kernel of $\ad_X$.  To prove the
forward implication, suppose ${\cal P}(Z) = Z$.  Then $Z = \sum_j P_j
Z P_j$.  Thus $\ad_X(Z) = \sum_j [X,P_j Z P_j] = 0$, as required.  To
prove the reverse implication, suppose $\ad_X(Z) = 0$.  Then
$[X,Z]=0$, and we conclude that it must be possible to write $Z =
\sum_j Z_j$, where $Z_j$ acts only within the eigenspace corresponding
to the eigenvalue $x_j$.  It follows that ${\cal P}(Z_j) = Z_j$, and
thus ${\cal P}(Z) = Z$, as desired.
\qed

To compute ${\cal E}_X(Z)$ we first consider two special cases: the
case when $Z$ is an element of $\mbox{ker}(\ad_X)$, and the case when
$Z$ is in the orthocomplement to $\mbox{ker}(\ad_X)$.

\textbf{Case: $Z \in \mbox{ker}(\ad_X)$}. We see from inspection of
the power series expansion Equation~(\ref{eq:power-series}) that all but
the first term vanishes, leaving ${\cal E}_X(Z) = Z$.

\textbf{Case: $Z$ in the orthocomplement to $\mbox{ker}(\ad_X)$}.  In
this space $\ad_X$ has a Moore-Penrose generalized inverse.  It is
most convenient to express ${\cal E}_X(Z)$ in vectorized form (c.f.
Equation~(\ref{eq:app-vec-Ex}))
\begin{eqnarray}
  \mbox{vec}({\cal E}_X(Z))
  & = & i (U^* \otimes U-I\otimes I) (X^*\otimes I-I \otimes X)^{-1} \times
  \nonumber \\
  & & \mbox{vec}(Z),
\end{eqnarray}
where the inverse operation is the Moore-Penrose generalized inverse,
easily computed by any of the standard computer algebra packages.
  
\textbf{General case:} We can now compute a general expression for
${\cal E}_X(Z)$ by combining these two special cases and our
expression for ${\cal P}$, the projector onto $\mbox{ker}(\ad_X)$.  We
have
\begin{eqnarray}
  \mbox{vec}({\cal E}_X(Z))
  & = & \mbox{vec}({\cal E}_X({\cal P}(Z))) +
  \mbox{vec}({\cal E}_X({\cal Q}(Z))), \nonumber \\
  & & 
\end{eqnarray}
where ${\cal Q} \equiv {\cal I}-{\cal P}$ projects onto the
orthocomplement of $\mbox{ker}(\ad_X)$.  Using the previous
observations we have
\begin{eqnarray} 
  \mbox{vec}({\cal E}_X(Z))
  & = & \mbox{vec}({\cal P}) \mbox{vec}(Z)+ \nonumber \\
  & &
  i (U^* \otimes U-I\otimes I) (X^*\otimes I-I \otimes X)^{-1} \times
  \nonumber 
  \\
  & & 
  (I \otimes I - \mbox{vec}({\cal P})) 
\mbox{vec}(Z), \label{eq:explicit-evaluation}
\end{eqnarray}
where the inverse is again the Moore-Penrose generalized inverse.
Note also that we have
\begin{eqnarray} \label{eq:projector-expression}
  \mbox{vec}({\cal P}) = \sum_j P_j^T \otimes P_j,
\end{eqnarray}
where the $P_j$ project onto the eigenspaces of $X$ with distinct
eigenvalues.

Equations~(\ref{eq:explicit-evaluation})
and~(\ref{eq:projector-expression}) offer an explicit way of computing
the action of ${\cal E}_X$, and thus of making the change of variables
from natural Pauli co-ordinates to natural adapted co-ordinates on
$T_U SU(2^n)$.  In practice, of course, this computation may be rather
cumbersome, however it is in principle possible using the approach we
have described.

It is easy to invert Equation~(\ref{eq:explicit-evaluation}), obtaining
\begin{eqnarray} 
  & & \mbox{vec}({\cal E}^{-1}_X(Z)) \nonumber \\
  & = & \mbox{vec}({\cal P}) \mbox{vec}(Z) \nonumber \\
  & &
  -i (X^*\otimes I-I \otimes X) (U^* \otimes U-I\otimes I)^{-1} \times
  \nonumber 
  \\
  & & 
  (I \otimes I - \mbox{vec}({\cal P})) \mbox{vec}(Z),
\end{eqnarray}
where the inverse operation is a Moore-Penrose generalized inverse.
Using this expression we can explicitly compute the change of
variables from natural adapted co-ordinates to natural Pauli
co-ordinates on $T_U SU(2^n)$.

\section{The Pauli geodesics}
\label{sec:geodesic_properties}

In this section we'll study a class of curves which are geodesics for
each of our families of Finsler metrics, $F_{1\Delta}, F_{p\Delta},
F_2$, and $F_q$.  Our study begins in
Subsection~\ref{subsec:symmetries} where we identify some isometries
of our Finsler metrics.  In Subsection~\ref{subsec:Pauli} we use these
isometries and the geodesic equation to identify a special class of
geodesic solutions, which we call \emph{Pauli geodesics}.  These
solutions are geodesics for \emph{all} the local metrics we have
defined, although their lengths may be different for the different
local metrics.  Examining these Pauli geodesics, we find examples of
unitary operators with multiple (indeed, infinitely many) Pauli
geodesics passing through them.  In Subsection~\ref{subsec:CVP} we
show that the problem of determining the minimal length Pauli geodesic
passing between $I$ and a unitary operation $U$ which is diagonal in
the computational basis is equivalent to solving an instance of the
closest vector in a lattice problem (CVP).  Also in this subsection,
we show that if the minimal length curve from $I$ through $U$ is
\emph{unique} then it must be a Pauli geodesic, and so the length of
the minimal Pauli geodesic will be $d_F(I,U)$.
Subsection~\ref{subsec:vol-argument} uses the connection to CVP to
argue that all but a tiny fraction of unitaries diagonal in the
computational basis have exponentially long minimal Pauli geodesics.
The section concludes in
Subsection~\ref{subsec:discussion-of-Pauli-geodesics} with a
discussion of the results obtained, and some caveats about their
implications.

\subsection{Metric isometries}
\label{subsec:symmetries}

In order to understand the space of solutions to the geodesic
equation, it is helpful to first study isometries of the local metric,
$F$, which in turn are reflected in symmetry properties of the
geodesics.

What do we mean by an isometry of $F$?  Suppose $h : M \rightarrow M$
is a diffeomorphism of the Finsler manifold $M$ to itself. If $F$ is a
Finsler metric on $M$, then we say $h$ is an \emph{$F$-isometry} if
$l_F(s) = l_F(h \circ s)$ for all curves $s$.  It is clear that such
an isometry preserves geodesics on the manifold $M$, i.e. $h \circ s$
is a geodesic if and only if $s$ is a geodesic.  It is also
straightforward to see that a necessary and sufficient condition for
$h$ to be an isometry is that
\begin{eqnarray}
  F(s(t),[s]_t) = F((h \circ s)(t),[h \circ s]_t)
\end{eqnarray}
for all curves $s$ and for all $t$.  Note that $[h \circ s]_t =
h_*[s]_t$, where $h_* : T_{s(t)}M \rightarrow T_{h(s(t))} M$ is the
linear pushforward map, so this condition may be rewritten:
\begin{eqnarray} \label{eq:symmetry-condition}
  F(x,y) = F(h(x),h_*y)
\end{eqnarray}
for all $(x,y) \in TM$.

For a local metric $F : T \, SU(2^n) \rightarrow [0,\infty)$ the
condition Equation~(\ref{eq:symmetry-condition}) that $h$ be an
isometry is equivalent to the condition
\begin{eqnarray} 
  F(U,H) = F(h(U),h_* H),
\end{eqnarray}
where $h_*$ is a superoperator pushing forward the Hamiltonian $H$
representing the tangent at $U$.  When $F$ is right-invariant this may
be replaced by the condition
\begin{eqnarray} \label{eq:adapted-symmetry-condition}
  F(H) = F(h_* H),
\end{eqnarray}
where $h_*$ is (implicitly) a function of the location $U$ on
$SU(2^n)$, and Equation~(\ref{eq:adapted-symmetry-condition}) must be
true at all values of $U$.

It will be convenient to regard $h_*$ as a matrix written with respect
to the $\sigma$ basis.  For all our local metrics a sufficient
condition for Equation~(\ref{eq:adapted-symmetry-condition}) to hold
is that $h_*$ be diagonal with respect to this basis, with entries
$\pm 1$.  This corresponds to the condition that $F(H)$ does not
depend on the sign of the expansion coefficients in $H = \sum_\sigma
\gamma^\sigma \sigma$, but only on their absolute values.  We will
call any right-invariant local metric with this property a
\emph{Pauli-symmetric} local metric.  It is clear that
$F_{1\Delta},F_{p\Delta},F_2$ and $F_q$ are all Pauli-symmetric local
metrics.  Some of our local metrics admit larger classes of
isometries:
\begin{itemize}
\item $F_{1\Delta}$: It suffices that $h_*$ be a \emph{signed
    permutation}, i.e., there is a permutation $\pi$ of the Pauli
  matrices such that $h_*(\sigma) = \pm {\pi(\sigma)}$.
  
\item $F_{p \Delta}$: It suffices that $h_*$ be a block diagonal sum
  of signed permutations, where the blocks correspond to all those
  values of $\sigma$ for which $p(\mbox{wt}(\sigma))$ takes the same
  value.

\item $F_2$: It suffices that $h_*$ be an orthogonal matrix.
  
\item $F_{q}$: It suffices that $h_*$ be a block diagonal sum of
  orthogonal matrices, where the blocks correspond to all those values
  of $\sigma$ for which $q(\mbox{wt}(\sigma))$ takes the same value.

\end{itemize}
These classes of isometry sometimes impose severe constraints on the
form of $h$.  For example, the continuity of $h$ and the fact that
$SU(2^n)$ is connected imply that if $h_*$ is a signed permutation for
all values of $U$, then $h_*$ must be a constant.  It is not difficult
to prove that this constant uniquely determines $h_*$, so the set of
such $h_*$ can be labeled by the signed permutations, of which there
are only a finite number.  Indeed, it is possible that the class of
isometries may be even further constrained: it is not \emph{a priori}
clear that given a particular signed permutation there even exists an
$h$ such that $h_*$ takes on the value of that signed permutation
everywhere.

The problem of obtaining a complete classification of the isometries
is an interesting problem in its own right, but it is not our main
concern here.  Rather, we will construct some explicit examples of
isometries $h$ realizing one or more of these conditions, and use
those isometries to construct the Pauli geodesics.

\textbf{Example: adjoint action of the Pauli group.}  Suppose $\sigma$
is a generalized Pauli matrix.  We can define a corresponding map
$h_\sigma : SU(2^n) \rightarrow SU(2^n)$ by $h_\sigma(U) \equiv \sigma
U \sigma^\dagger$.  A straightforward calculation shows that
$h_{\sigma*}(H) = \sigma H \sigma^\dagger$, so $h_{\sigma *}$ is
indeed diagonal with entries $\pm 1$, and $h_\sigma$ is an isometry of
all our local metrics.

\textbf{Example: complex conjugation.}  Define the map $h : SU(2^n)
\rightarrow SU(2^n)$ by $h(U) \equiv U^*$.  A calculation shows that
$h_*(H) = -H^*$, and thus $h_*$ is again diagonal with entries $\pm
1$.  It follows that the map $U \rightarrow U^*$ is an isometry of all
our local metrics.

\textbf{Example: adjoint action of local unitaries.} Suppoe $W \equiv
W_1 \otimes \ldots \otimes W_n$ is a local unitary operation on $n$
qubits.  Define $h_W: SU(2^n)\rightarrow SU(2^n)$ by $h_W(U) \equiv W
U W^\dagger$.  Then $h_*(H) = W H W^\dagger$, whence $h_W$ is an
isometry for $F_2$ and $F_q$.

\textbf{Example: adjoint action of the Clifford group.}  Recall that
the $n$-qubit Clifford group\footnote{Sometimes referred to as the
  \emph{normalizer} of the Pauli group.  See Chapter~10 of
  \cite{Nielsen00a} for a review of the Clifford group and the
  associated stabilizer formalism, or~\cite{Gottesman97a} for much of
  the original development of this formalism and its applications in
  quantum information science.}  consists of all those $n$-qubit
unitary operations $g$ having the property that $g \sigma g^\dagger$
is a generalized Pauli matrix whenever $\sigma$ is a generalized Pauli
matrix.  This group includes many interesting unitary operations,
including the controlled-{\sc not}, the Hadamard gate, and the
generalized Pauli matrices themselves.

Suppose $g$ is an element of the Clifford group.  We can define a
corresponding map $h_g : SU(2^n) \rightarrow SU(2^n)$ via $h_g(U)
\equiv g U g^\dagger$.  We compute the pushforward $h_{g*}$ at $U$,
obtaining $h_{g*}(H) = g H g^\dagger$.  Since $g$ is an element of the
Clifford group, it follows that $h_{g*}$ is a permutation with respect
to the $\sigma$ co-ordinates, and thus $h_g$ is an isometry of $F_{1
  \Delta}$ and of $F_2$, but not in general of $F_{p \Delta}$ or of
$F_q$.

\textbf{Example: adjoint action of the unitary group on $SU(2^n)$.}
Let $W$ be an arbitrary unitary, and define an action $h_W : SU(2^n)
\rightarrow SU(2^n)$ by $h_W(U) \equiv W U W^\dagger$.  A calculation
shows that $h_{W*}(H) = W H W^\dagger$, and thus $h_W$ is an isometry
of $F_2$, but is not in general an isometry of the other local
metrics.

\subsection{Pauli geodesics}
\label{subsec:Pauli}

The isometries identified in the previous subsection enable us to
identify a large class of geodesics which we call \emph{Pauli
  geodesics}.  These geodesics arise as a result of the Pauli group
isometry, and thus are geodesics for all the local metrics we have
defined, and, indeed, of any Pauli-symmetric local metric.

To construct the Pauli geodesics we begin with a simple proposition.

\begin{proposition} {} \label{prop:symmetry}
  Let $M$ be a Finsler manifold.  Suppose $h : M \rightarrow M$ is an
  isometry and $s$ is a geodesic such that (a) $h(s(0)) = s(0)$, and
  (b) $h_*([s]_0) = [s]_0$.  Then $s = h \circ s$ and $h_*([s]_t) =
  [s]_t$ for all $t$.
\end{proposition}

\textbf{Proof:} The proof is to observe that $s$ and $h \circ s$ are
both geodesics with the same starting point, $h(s(0)) = s(0)$, and the
same initial tangent vector $h_*([s]_0) = [s]_0$.  The geodesic
equation is a second order ordinary differential equation, and thus by
the uniqueness of solutions to such equations we deduce that $h \circ
s = s$.  It follows immediately that $h_*([s]_t) = [s]_t$ for all $t$.
\qed

As a simple but useful illustration of the proposition, suppose we
have a solution $U(t)$ to the geodesic equation for a Pauli-symmetric
Finsler metric, with initial tangent vector corresponding to a
Hamiltonian $H_0$.  Suppose $\sigma H_0 \sigma^\dagger = H_0$ for some
generalized Pauli matrix $\sigma$.  Then
Proposition~\ref{prop:symmetry} implies that $\sigma U(t)
\sigma^\dagger = U(t)$ for all $t$.

The construction of the Pauli geodesics is based on the stabilizer
formalism developed by Gottesman~\cite{Gottesman97a}; for a review,
see Chapter~10 of~\cite{Nielsen00a}.  In particular, we suppose
$\sigma_1,\ldots,\sigma_n$ is a set of stabilizer generators, i.e.,
independent and commuting generalized Pauli matrices which generate a
subgroup $S$ of the full Pauli group.  Suppose $H_0 = \sum_{\sigma \in
  S} h^\sigma \sigma$.  Then we claim that the geodesic $U$ with $U(0)
= I$ and initial tangent vector corresponding to $H_0$ is just $U(t) =
\exp(-i H_0 t)$, for any Pauli-symmetric Finsler metric.  We refer to
$U(t)$ as a Pauli geodesic for the Finsler metric.

The first step of the proof is to observe that the Pauli co-ordinates
$x^\sigma(t)$ of $U(t)$ are identically zero, unless $\sigma \in S$.
To see this, suppose $\sigma \not \in S$, and choose $\tilde \sigma
\in S$ which anticommutes with $\sigma$.  From our earlier remarks we
see that $U(t) = \tilde \sigma U(t) \tilde \sigma^\dagger$, and thus
$x^\sigma(t) = 0 $ for all $t$.  We say that a unitary satisfying this
condition is \emph{$S$-invariant}.

We now analyse the geodesic equation,
Equation~(\ref{eq:Euler-Lagrange-2}), for the co-ordinates $x^\sigma$
and $y^\sigma$ with $\sigma \in S$.  In particular, because $U(t)$ is
guaranteed to be $S$-invariant we may effectively regard $F^2$ as a
function of $x^\sigma$ and $y^\sigma$ only for $\sigma \in S$.  We
have
\begin{eqnarray} \label{eq:Pauli-geo}
  \frac{d}{dt} \left( \frac{\partial F^2}{\partial y^\sigma} \right) =
  \frac{\partial F^2}{\partial x^\sigma},
\end{eqnarray}
where all partial derivatives are evaluated at $S$-invariant unitary
matrices.  But $\partial F^2 / \partial x^\sigma = 0$ at such an
$S$-invariant unitary matrix, since $F^2$ has no dependence on
$x^\sigma$, by the commutativity of the elements of $S$.  Substituting
this into the right-hand side of Equation~(\ref{eq:Pauli-geo}) and
applying the chain rule to the left we obtain:
\begin{eqnarray}
  \sum_{\tau \in S} \left(
  \frac{\partial F^2}{\partial x^\tau \partial y^\sigma} 
  \frac{dx^\tau}{dt} +
  \frac{\partial F^2}{\partial y^\tau \partial y^\sigma} 
  \frac{dy^\tau}{dt} \right)
  = 0.
\end{eqnarray}
But $\frac{\partial F^2}{\partial x^\tau \partial y^\sigma} = 0$,
since $\frac{\partial F^2}{\partial x^\tau} = 0$, and thus
\begin{eqnarray}
  \sum_{\tau \in S}
  \frac{\partial F^2}{\partial y^\tau \partial y^\sigma} 
  \frac{dy^\tau}{dt}
  = 0.
\end{eqnarray}
Using the invertibility of the Hessian we obtain $\frac{dy^\tau}{dt} =
0$, and thus $x^\tau = c^\tau t$ for some constant $c^\tau$.  It
follows that the solution to the geodesic equation is
\begin{eqnarray}
  U(t) = \exp(-i H_0 t),
\end{eqnarray}
as claimed.

Summing up, for a Pauli-symmetric Finsler metric such as $F_{1\Delta},
F_{p\Delta}, F_2$ or $F_q$, when the initial Hamiltonian $H_0$ is a
sum over terms in a stabilizer subgroup, the corresponding Pauli
geodesic solution is just the exponential $U(t) = \exp(-i H_0 t)$.

\subsection{Minimal Pauli geodesics and the closest vector in a lattice 
problem}
\label{subsec:CVP}

In this subsection we study the minimal length Pauli geodesics from
$I$ to $U$, where $U$ is diagonal in the computational basis.  We show
that for any right-invariant Pauli-symmetric Finsler metric this
minimal length is equal to the solution of an instance of the closest
vector in a lattice problem (CVP).  This class of Finsler metrics
include all the Finsler metrics of most interest to us: $F_{1\Delta},
F_{p\Delta}, F_2$, and $F_q$.

Note that the case where $U$ is diagonal in the computational basis
corresponds to the case where the stabilizer $S$ contains exactly the
products of Pauli $I$ and $Z$ matrices, e.g., for $n = 2$, $S$
contains $II, ZI, IZ$ and $ZZ$.  Exactly analogous results hold for
all other choices of stabilizer, but working with this particular
stabilizer allows us to make use of certain standard notations and
nomenclature, and so avoid the introduction of extra terminology.

One reason for specializing to unitaries diagonal in the computational
basis is that it includes a class of unitaries of exceptional
interest: those that can be written $U_f = \sum_z (-1)^{f(z)}
|z\rangle \langle z|$, where $f(z)$ is a classical Boolean function on
the $n$-bit input $z$.  Kitaev's phase estimation
algorithm~\cite{Kitaev97b} shows that, given a single ancilla qubit,
the computation of $U_f$ requires essentially the same number of
quantum gates as computation of the function $f$ on a quantum
computer.  Thus, bounds on the size of the circuit required to compute
$U_f$ are of considerable interest.

Returning to the general case of $U$ diagonal in the computational
basis, our goal in this subsection is to study the length of the
minimal Pauli geodesic between $I$ and $U$.  Of course, the quantity
of real interest to us is the length of the minimal geodesic between
$I$ and $U$, unconstrained by the constraint that it be a Pauli
geodesic.  Unfortunately, we can't say when it will be true that the
minimal length geodesic is going to be a Pauli geodesic.  However, the
following proposition gives some hopes that this will be the case for
some unitaries of interest.

\begin{proposition}{} \label{prop:uniqueness}
  Let $F$ be a Pauli-symmetric Finsler metric.  Let $U \in SU(2^n)$ be
  diagonal in the computational basis.  Suppose the minimal length
  geodesic $s$ between $I$ and $U$ is unique, i.e., there is only a
  single curve $s$ between $I$ and $U$ with $d_F(I,U) = l_F(s)$.  Then
  $s$ must be a Pauli geodesic.
\end{proposition}

For the usual model spaces of Riemannian geometry (the sphere, flat
Euclidean space, or the hyperbolic space) non-unique minimal paths are
quite non-generic, suggesting that the same may be true in the
situations of interest to us\footnote{Compare, however, the
  counterexample in
  Subsection~\ref{subsec:discussion-of-Pauli-geodesics}, below.}.

\textbf{Proof:} Let $\sigma$ be a generalized Pauli matrix containing
only $Z$s and $I$s.  Let $s$ be the minimal length geodesic between
$I$ and $U$.  Define $\tilde s(t) \equiv \sigma s(t) \sigma^\dagger$.
Then $\tilde s$ has the same endpoints and length as $s$, and thus, by
the uniqueness of the minimal geodesic, we must have $\tilde s = s$.
Since this is true for all $\sigma$ containing only $Z$s and $I$s, it
follows that $s(t)$ is diagonal in the computational basis for all
$t$, and thus $s$ is a Pauli geodesic.
\qed

Fixing $U$, which Pauli geodesics pass from $I$ to $U$?  To answer
this question, choose Hermitian $H$ such that $U = \exp(-i H)$.  Let
${\cal J}$ be the set of traceless Hermitian matrices which are
diagonal in the computational basis, and have diagonal entries which
are integer multiples of $2 \pi$.  Let $J \in {\cal J}$.  Then for any
such $J$, the curve $\exp(-i(H-J)t)$ is a Pauli geodesic passing
through $U$.

This freedom to choose $J$ actually exhausts the freedom in the choice
of Pauli geodesics\footnote{Note that in analysing the freedom we
  restrict ourselves to Pauli geodesics $\exp(-iH't)$ for which $H'$
  is diagonal in the computational basis.  For some very non-generic
  $U$ it may be that $U$ is diagonal in the computational basis, yet
  has a Pauli geodesic passing through it for which $H'$ is not
  diagonal in the computational basis.  We shall ignore this
  possibility.}  passing through $U$.  To see this, suppose $\exp(-i H
t)$ and $\exp(-i H't)$ are two Pauli geodesics passing through $U$ at
$t = 1$.  Then we have $\exp(-i H) = \exp(-i H')$, whence
$\exp(i(H'-H)) = I$, since $H$ and $H'$ are both diagonal in the
computational basis, and thus commute.  However, in order that
$\exp(i(H'-H)) = I$, we must have that $J \equiv H-H'$ is traceless
(since both $H$ and $H'$ are), and diagonal in the computational
basis, with entries which are integer multiples of $2\pi$.

It is straightforward to verify that the set ${\cal J}$ has the
structure of a lattice, i.e., taking an integer linear combination of
elements of ${\cal J}$ produces another element of ${\cal J}$.  A
basis for this lattice is the matrices $2\pi(|z\rangle \langle
z|-|0\rangle \langle 0|)$, where $z \neq 0$.

The length of the Pauli geodesic $\exp(-i(H-J)t)$ between $I$ and $U$
is given by $F(H-J)$, so the length of the minimal Pauli geodesic
through $U$ is given by:
\begin{eqnarray}
  \min_{J \in {\cal J}} F(H-J).
\end{eqnarray}
Thus the problem of finding the minimal length Pauli geodesic is
equivalent to finding the lattice vector in ${\cal J}$ closest to $H$
according to the $F(\cdot)$ norm on $su(2^n)$.  This is the desired
connection to the closest vector in a lattice problem.

The connection to lattices also makes it straightforward to construct
arbitrarily long geodesics passing through a given unitary.  This is
true, for example, even in the two-qubit case.  Suppose we choose $U =
\exp(-i \pi ZZ / 2)$, and select $H = \pi ZZ / 2+ 2 \pi ZI / M$, where
$M$ is a positive integer equal to $1$ modulo $4$.  Then $\exp(-i Ht)$
is a Pauli geodesic which first passes through $U$ at $t = M$.  By
making $M$ sufficiently large we can increase the length of this
geodesic without bound.

\subsection{Existence of exponentially long minimal Pauli geodesics}
\label{subsec:vol-argument}

In the previous subsection we showed that finding the minimal length
Pauli geodesic from $I$ through a diagonal unitary $U$ is equivalent
to solving an instance of CVP.  In this subsection we'll use this
connection to prove that for most such $U$ the minimal length Pauli
geodesic is exponential in length.  The key is the following
proposition, pointed out to the author by Oded Regev:

\begin{proposition}{} \label{prop:vol-argument}
  Let $V$ be a $d$-dimensional vector space equipped with the standard
  Lebesgue measure.  Let $F$ be a norm on $V$, and let $V_F(r)$ be the
  Lebesgue measure of the unit ball of radius $r$ associated to $F$.
  Let ${\cal J}$ be a $d$-dimensional lattice in $V$, and let $M$ be a
  matrix whose columns are a lattice basis for ${\cal J}$, so the
  Lebesgue measure of a unit cell in ${\cal J}$ is $\det(M)$.  Then if
  a fraction $f$ $(0 \leq f \leq 1)$ of points in $V$ are within a
  distance $r$ of the lattice we must have
  \begin{eqnarray}
    f \det(M) \leq V_F(r).
  \end{eqnarray}
\end{proposition}

\textbf{Proof:} Consider a large volume containing $N$ lattice points.
Let $R$ be the region obtained by surrounding each of the $N$ lattice
points by the unit ball of radius $r$ according to the norm $F$.  The
total Lebesgue measure of the region $R$ is at most $N V_F(r)$.  By
assumption, in the large $N$ limit $R$ contains at least a fraction
$f$ of points in the $N$ unit cells associated to the $N$ lattice
points\footnote{We neglect finite-size corrections of order sublinear
  in $N$.}, and thus $f N \det(M) \leq N V_F(r)$.  Dividing by $N$
gives the desired result. \qed

To apply this result, it simplifies matters\footnote{Analogous results
  hold for $SU(2^n)$, but the calculations are more complicated, due
  to the more complex lattice basis.} to vary our earlier approach
slightly, moving from $SU(2^n)$ to $U(2^n)$, and defining the local
metrics $F_1,F_{1\Delta},F_p,F_{p\Delta},F_2,F_q$ analogously to
before, but now with a contribution from the $\sigma = I^{\otimes n}$
term.  It is not difficult to show that when $U$ is in $SU(2^n)$ it
has the same minimal curves regardless of whether we use the
formulation of the local metric in $SU(2^n)$ or $U(2^n)$.

In this formulation, Pauli geodesics exist for all our Finsler
metrics, and the minimal length Pauli geodesic is found by minimizing
$F(H-J)$, where $J$ is in the lattice spanned by matrices of the form
$2\pi |z\rangle \langle z|$, which may be rewritten in the more
convenient form $2\pi \otimes_{j=1}^n (I+z_j Z_j)/2^n$.  Arranged into
columns, the corresponding matrix of lattice basis vectors has the
form $2\pi H^{\otimes n}/2^{n/2}$, where $H$ is the usual $2 \times 2$
Hadamard matrix.  Thus the conclusion of
Proposition~\ref{prop:vol-argument} is that
\begin{eqnarray} \label{eq:inter-vol}
  f \times \left( \frac{2\pi}{2^{n/2}}\right)^{2^n} \leq V_F(r).
\end{eqnarray}
Let us analyse what this allows us to conclude about the fraction $f$
of points within a distance $r$ of the lattice, for each of our
choices of Finsler metric.

\textbf{Case: $F_{1\Delta}$.} As $\Delta \rightarrow 0$ the unit
sphere has volume $V_{F_{1\Delta}}(r) \rightarrow (2r)^{2^n}/(2^n!)$.
Applying Stirling's formula, Equation~(\ref{eq:inter-vol}) reduces to
\begin{eqnarray}
  r \geq \frac{\pi}{e} 2^{n/2} f^{1/2^n}
\end{eqnarray}
in the $\Delta \rightarrow 0$ limit.  In consequence, unless $r$ is
exponentially large, at most a doubly exponentially small fraction of
diagonal unitary operators will have minimal Pauli geodesics of length
$r$ or less.

\textbf{Case: $F_{p\Delta}$.} Obviously, the minimal length Pauli
geodesics for $F_{p\Delta}$ are longer than those for $F_{1\Delta}$,
provided the penalty function satisfies $p(j) \geq 1$.  Thus, unless
$r$ is allowed to be exponentially large, at most a doubly
exponentially small fraction of diagonal unitaries will have minimal
Pauli geodesics of length $r$ or less.

\textbf{Case: $F_2$.}  Based on our previous results we expect
constant size minimal length Pauli geodesics for $F_2$.  This
expectation is not disappointed.  The volume formula in this case is
$V_{F_2}(r) = (\sqrt{\pi} r)^{2^n}/(2^n/2)!$.  Applying Stirling's
formula, and setting $f=1$, Equation~(\ref{eq:inter-vol}) reduces to:
\begin{eqnarray}
  \sqrt{\frac{2\pi}{e}} \leq r.
\end{eqnarray}

\textbf{Case: $F_q$.}  The volume elements is given by
\begin{eqnarray}
  V_{F_q} = \frac{\left( \sqrt{\pi} r\right)^{2^n}}{\left(2^n/2\right)!}
\prod_\sigma \frac{1}{q(\mbox{wt}(\sigma))},
\end{eqnarray}
where the product is taken over all $\sigma$ containing only $I$ and
$Z$ terms.  Applying Stirling's formula, Equation~(\ref{eq:inter-vol})
reduces to:
\begin{eqnarray}
  f^{1/2^n} \sqrt{\frac{2\pi}{e}}
  \left( \prod_\sigma q(\mbox{wt}(\sigma)) \right)^{1/2^n} \leq r.
\end{eqnarray}
We see that provided the penalty function $q$ is chosen appropriately,
all but a doubly exponentially small fraction of diagonal unitaries
will have minimal Pauli geodesics which are exponentially long.  Such
a choice is provided, for example, by
Equation~(\ref{eq:example-penalty}), with $k$ exponentially large.

\subsection{Discussion}
\label{subsec:discussion-of-Pauli-geodesics}

In the past few subsections we've explained the construction of the
Pauli geodesics, connected the minimal length Pauli geodesic to the
solution of an instance of CVP, shown that the minimal length Pauli
geodesic is actually the minimal length curve, provided that curve is
unique, and proved that most diagonal unitaries have exponential
length minimal Pauli geodesics.  This subsection injects some words of
warning into this otherwise encouraging situation, explaining some
significant caveats to our results.

\subsubsection{On the uniqueness of minimal curves} 

Based on the standard model spaces of Riemannian geometry (the sphere,
Euclidean flat space, or the hyperbolic space), it seems plausible
that the minimal geodesics between $I$ and a diagonal unitary $U$ are
generically unique, and thus Pauli geodesics.  However, this may not
always be the case for $F_p$, as the following example shows.

Consider the Boolean function $f(z) = f(z_1,\ldots,z_n) \equiv z_1 z_2
\ldots z_n$, i.e., the {\sc AND} of the $n$ bits $z_1,\ldots,z_n$, and
the associated unitary transformation $U|z\rangle \equiv
(-1)^{f(z)}|z\rangle$.  Using the connection to CVP, it is easy to
verify that for all of our Finsler metrics $F$ the minimal length
Pauli geodesic is $\pi F(|1,\ldots,1\rangle \langle 1,\ldots,1|)$. 

Consider the case of $F_{p\Delta}$, with the penalty function $p$
chosen as $q$ was in Equation~(\ref{eq:example-penalty}).  A
calculation shows that as $\Delta \rightarrow 0$ the minimal Pauli
geodesic has length:
\begin{eqnarray} \label{eq:min-Pauli-and}
  \pi \left(k - \frac{2+n+n^2}{2^{n+1}}(k-1)\right).
\end{eqnarray}
When $k$ is large, this is dominated by the term $\pi k$.

By contrast, the results of Barenco \emph{et al}~\cite{Barenco95a}
show that there is a quantum circuit for $U$ containing $O(n^2)$ one-
and two-qubit gates.  It follows that $d_{F_p}(I,U) \leq c n^2$, for
some constant $c$.  To reconcile this result with
Equation~(\ref{eq:min-Pauli-and}) we see that when $k$ is large, the
minimal geodesic between $I$ and $U$ \emph{must not} be a Pauli
geodesic, and therefore cannot be unique.  This example --- which is
easily modified to give other examples --- suggests that the
applicability of Proposition~\ref{prop:uniqueness} may be limited, at
least for some choices of Finsler metric, and highlights the need to
develop more tools for the analysis of the distance function
$d_F(I,U)$.

\subsubsection{Classical simulations} 

We have argued earlier in the paper that it is at least plausible that
local metrics such as $F_1, F_p$ and $F_q$ give rise to distance
functions $d_F(I,U)$ which are polynomially equivalent to $m_{\cal
  G}(U)$.  Suppose, for the sake of argument, that we can find a
Finsler metric $F$ with this property.  Suppose furthermore that for a
generic unitary diagonal in the computational basis, the minimal
length curve is unique.  If this is the case, then for such unitaries
there is circuit containing only gates diagonal in the computational
basis, and with a size polynomially equivalent to the minimal number
of gates required to generate $U$.

This conclusion would be rather surprising, as such circuits can be
simulated with at most a polynomial overhead in the classical circuit
model, and it would therefore conflict with the general belief that
quantum computers offer a substantial complexity advantage over
classical computers.

Of course, there are many potential loopholes in this argument: it
makes use of many unstated assumptions (no use of ancillas, no
approximation, no uniformity requirement) as well as several steps
that, while plausible, could easily turn out to be wrong.  I can not
at present resolve which of the many possibilities is correct, but it
suggests many interesting directions for further research.

\section{Conclusion}
\label{sec:conclusion}

In this paper we have proposed a geometric approach to the problem of
proving lower bounds on the number of quantum gates required to
synthesize a desired unitary operation. In particular, we have shown
that such lower bounds may be provided by the length of the minimal
geodesics of certain Finsler metric structures on $SU(2^n)$.

Our main progress in understanding the geodesic structure for these
Finsler metrics are the results: (1) a method for computing the
geodesic equation explicitly, thus enabling numerical investigations;
(2) the construction of a large class of geodesic solutions, which we
call Pauli geodesics, passing from the identity $I$ through any
unitary $U$ which is diagonal in the computational basis; (3) the
demonstration of an equivalence between the problem of finding the
minimum length Pauli geodesic between $I$ and $U$, and the closest
vector in a lattice problem (CVP); (4) a proof that when there is a
\emph{unique} minimal length geodesic between $I$ and $U$, then that
geodesic must be a Pauli geodesic; and (5) a proof that all but a very
small fraction of diagonal unitaries $U$ have minimal length Pauli
geodesics which are of exponential length.

To make further progress it will be necessary to obtain more insight
into the space of geodesics associated to each of our Finsler metrics.
Of course, understanding the space of geodesics associated to a
Finsler metric is a difficult problem to solve, even for relatively
simple Riemannian spaces, and much of the ongoing work in Riemannian
and Finsler geometry is motivated by this problem.

Questions of particular interest include: (a) what are the geodesics;
(b) how long are the geodesics, and can we find the minimal length
geodesics, or at least a bound on their length; (c) do there exist
exponentially long minimal length geodesics, and if so, can we
construct some explicit examples, and hence explicit examples of
unitary operations requiring an exponential number of gates; and (d)
for which (if any) local metric $F$ is the minimal path length
$d_F(I,U)$ polynomially equivalent to the size $m_{\cal G}(U)$ of the
minimal quantum circuit synthesizing $U$?

Broadening the scope, the results of this paper do not yet address
many issues of interest in quantum computational complexity.  In
particular, our results are constrained entirely to \emph{exact} and
\emph{non-uniform} implementations of a unitary operation, while the
subject of most interest for quantum computational complexity is
approximate and uniform implementations.  Also from the point of view
of computational complexity, it is desirable to obtain strong results
about the impact working qubits (i.e., ancilla) have on minimal path
lengths.  Finally, it is tempting to speculate on whether a geometric
approach along the lines sketched here could ever be powerful enough
to resolve complexity class class separations.  In this vein, it
should be noted that results such as the well-known no-go theorem of
Razborov and Rudich~\cite{Razborov94a} (see also~\cite{Aaronson03a})
suggest that to apply the geometric approach to such separations would
require deep insights into very specific computational problems.

On the flip side, one might ask if these techniques can be of any use
in quantum algorithm design, either for recovering existing
algorithms, or perhaps in the design of new algorithms.  In
particular, if we can find a local metric $F$ such that $d_F(I,U)$ is
polynomially equivalent to $m_{\cal G}(U)$, then quantum circuit
design may be viewed in terms of the construction of short geodesics
between $I$ and the desired unitary operation, i.e., in terms of the
solution of a two-point boundary value problem.  In a similar vein,
application of these ieas to oracle problems and quantum communication
complexity may be possible.

Further afield still, one may ask whether a similar approach based on
Finsler geometry might be taken in the study of classical computing.
\emph{A priori} this idea does not appear particularly promising, as
classical computing models are usually formulated in a discrete
fashion not amenable to study using the calculus of variations.
However, if one reformulates those models using the theory of
continuous time, discrete state space Markov chains, I believe it may
be possible to apply similar techniques, perhaps along the lines which
have been explored in the theory of optimal stochastic control.

\appendix
\section{Approximating local metrics with Finsler metrics}
\label{app:approximating-metrics}

In this appendix we explain how the local metrics $F_1$ and $F_p$,
which lack the smoothness and strong convexity properties required by
Finsler metrics, can be approximated arbitrarily well by Finsler
metrics.

To begin, let's formalize the notion of approximating one local metric
by another.  Let $F, \tilde F : TM \rightarrow [0,\infty)$ be two
local metrics on the manifold $M$.  We say $F$ is \emph{metrically
  equivalent} to $\tilde F$ if there exist positive constants $A$ and
$B$ such that
\begin{eqnarray}
  A \, F(x,y) \leq \tilde F(x,y) \leq B \, F(x,y)
\end{eqnarray}
for all $(x,y) \in TM$.  A little thought shows that this definition,
which appears asymmetric in $F$ and $\tilde F$, is actually symmetric.
It is also easy to see that if $F$ and $\tilde F$ are metrically equivalent
then they satisfy
\begin{eqnarray} \label{eq:distance-approximation}
  A \, d_F(x_1,x_2) \leq d_{\tilde F}(x_1,x_2) \leq B \, d_F(x_1,x_2),
\end{eqnarray}
for all $x_1$ and $x_2 \in M$.  

Our goal in this appendix is to construct Finsler metrics
$F_{1\Delta}$ and $F_{p\Delta}$ which are metrically equivalent to
$F_1$ and $F_p$, respectively, for sufficiently small values of the
positive parameter $\Delta$.  Furthermore, as $\Delta \rightarrow 0$
it turns out that $A \rightarrow 1$ and $B \rightarrow 1$ for both
classes of metrics, so Equation~(\ref{eq:distance-approximation})
tells us that the notion of length given by $F_{1\Delta}$ and
$F_{p\Delta}$ approaches that given by $F_1$ and $F_p$.

Our strategy in constructing the approximating Finsler metrics is to
first study the problem of finding Minkowski norms $N_\Delta$ which
approximate a given norm\footnote{Note that in keeping with standard
  usage in Finsler geometry we only require norms to be positively
  homogeneous, not homogeneous, as is usually stipulated in other
  contexts.} $N$ on $\mathbb{R}^d$.  Once we understand this problem
it is straightforward to construct the appropriate Finsler metrics.

Constructing Minkowski norms with suitable properties does not seem
easy to do directly, in large part because of the positive homogeneity
condition for norms.  We will take a more indirect approach to the
definition, defining norms in terms of their indicatrices, i.e., their
unit spheres.  This material, described in
Subsection~\ref{subsec:indicatrix}, is well-known in the Finsler
geometry literature (see, e.g.~\cite{Alvarez98a}), and our discussion
merely outlines the major facts.  The exception is
Proposition~\ref{prop:strong-convexity}, which seems to be a
well-known folklore result, but which I have not found explicitly in
the prior literature.  Consequently, a proof is included.
Subsection~\ref{subsec:approximations-norm} uses this background to
construct the desired classes of approximating Finsler metrics.

\subsection{The indicatrix and the implicit definition of Minkowski norms}
\label{subsec:indicatrix}

The following proposition gives a convenient way of defining smooth
norms in terms of a function $g : \mathbb R^d \rightarrow \mathbb R$
which is \emph{not} necessarily positively homogeneous.  To state the
proposition we define $S_g$ to be the set of points $y$ such that
$g(y) = 1$.  We will use $g$ to construct a norm $N_g$ whose
indicatrix is $S_g$.

\begin{proposition}
  Suppose $g : \mathbb R^d \rightarrow \mathbb R$ is smooth, convex,
  satisfies $g(0) < 1$, and is such that $S_g$ is compact.  Then the
  function $N_g$ defined by the equations
  \begin{eqnarray} 
  \label{eq:imp-defn-norm-1}
  N_g(0) & = & 0; \\
  \label{eq:imp-defn-norm-2}
  g\left( \frac{ y}{N_g( y)} \right) & = & 1, \mbox{ provided }  y
  \neq 0.
  \end{eqnarray}
  exists, is uniquely defined, is smooth away from the origin, and is
  a norm with indicatrix $S_g$.
\end{proposition}

\textbf{Proof:} This is easily proved, and a well-known result of
Finsler geometry.  See, for example,~\cite{Alvarez98a} for an outline
of the proof. The only non-trivial step is an application of the
implicit function theorem (see, e.g., Chapter~7 of~\cite{Lee03a}) to
the implicit definition~(\ref{eq:imp-defn-norm-2}) of $N_g$, in order
to obtain the smoothness condition for $N_g$.  \qed

When does $N_g$ obey the strong convexity constraint?  The following
proposition gives a simple criterion for $N_g$ to be strongly convex.

\begin{proposition} {} \label{prop:strong-convexity}
  Suppose $g: \mathbb R^d \rightarrow \mathbb R$ is such that the
  matrix $G$ with entries $\left( \frac{\partial g}{\partial y^j
      \partial y^k} \right)$ is strictly positive for any $ y \neq 0$.
  Then the norm $N_g$ defined by Equations~(\ref{eq:imp-defn-norm-1})
  and~(\ref{eq:imp-defn-norm-2}) is strongly convex.
\end{proposition}

\textbf{Proof:} To simplify notation we write $N = N_g$.  We begin by
differentiating the implicit definition Equation~(\ref{eq:imp-defn-norm-2})
with respect to $y^j$, obtaining
\begin{eqnarray}
  N_{,j}( y)
  = \frac{N( y) g_{,j}(\hat y)}
  {(\nabla_{\hat y} g \cdot  y)},
\end{eqnarray}
where $_{,j}$ denotes partial differentiation with respect to $y^j$,
$\hat y \equiv  y / N( y)$, and $\nabla_{\hat y}$ denotes the
usual gradient operator, evaluated at $\hat y$.  Differentiating
again, we obtain:
\begin{eqnarray}
  N_{,jk}( y)
  & = & \frac{ g_{,jk}(\hat y)}{\nabla_{\hat y} g \cdot  y} 
  + \frac{ g_{,j}(\hat y) g_{,k}(\hat y) \sum_{lm}
    g_{,lm}(\hat y) y^l y^m}{(\nabla_{\hat y} g \cdot  y)^3}
  \nonumber \\
  & & -\frac{\sum_l (g_{,j}(\hat y) g_{,kl}(\hat y) + 
    g_{,k}(\hat y) g_{,jl}(\hat y))y^l}
  {(\nabla_{\hat y} g \cdot  y)^2}. 
\end{eqnarray}
We also obtain
\begin{eqnarray}
  N^2_{,jk}( y)  = 2N( y) N_{,jk}( y) +
  2N_{,j}( y)N_{,k}( y).
\end{eqnarray}
Combining these results we obtain the Hessian:
\begin{eqnarray} \label{eq:Hessian}
  & & H_{jk}( y) \nonumber \\
  & = & \frac{ N g_{,jk}}{\nabla g \cdot  y} 
   + \frac{ N g_{,j} g_{,k}}
    {(\nabla g \cdot  y)^3} \left(
 \sum_{lm} g_{,lm} y^l y^m + N 
 \nabla g \cdot  y \right)
  \nonumber \\
  & & -\frac{N \sum_l (g_{,j} g_{,kl}+ 
    g_{,k} g_{,jl})y^l}
  {(\nabla g \cdot  y)^2},
\end{eqnarray}
where, to simplify notation, it is implicit that all derivatives of $g$
are evaluated at $\hat y$, and $N$ is evaluated at $ y$.

Examining Equation~(\ref{eq:Hessian}), we see that the contribution
from the first term on the right-hand side is strictly positive, since
both $N$ and $\nabla g \cdot y$ are strictly positive, and $g_{,jk}$
is strictly positive, by assumption.  The contribution from the second
term is positive, since it is a positive scalar multiple of the
positive matrix with components $g_{,j} g_{,k}$.  Thus the sum of the
first two terms is strictly positive.  The final term of
Equation~(\ref{eq:Hessian}) is more problematic, due to the presence
of the minus sign.

The resolution is to make a linear change of variables which
simplifies the Hessian.  In particular, we make a linear change of
variables so that $ y = (\alpha,0,0,\ldots,0)$, and $\nabla g =
(\beta,0,0,\ldots,0)$.  It is not difficult to see that such a linear
change of variables is always possible, and moreover does not affect
whether or not the Hessian is strictly positive.  It does, however,
make the analysis simpler.  In particular, observe that by homogeneity
we have:
\begin{eqnarray}
N^2(\alpha,0,\ldots,0) & = & \alpha^2 N(1,0,\ldots,0) \\ 
N^2_{,1}(\alpha,0,\ldots,0) & = & 2 \alpha N(1,0,\ldots,0) \\ 
N^2_{,11}(\alpha,0,\ldots,0) & = & 2 N(1,0,\ldots,0).
\end{eqnarray}
Observe also that for $j =2,\ldots, d$ we have
$N^2_{,j}(\alpha,0,\ldots,0) = 0$, since $\nabla_{ y} g =
(\beta,0,\ldots,0)$.  The homogeneity of $N$ then implies that
$N^2_{,j}(\alpha,0,\ldots,0) = 0$ for all $\alpha$, and thus
\begin{eqnarray}
  N^2_{,1j}(\alpha,0,\ldots,0) = N^2_{,j1}(\alpha,0,\ldots,0) = 0.
\end{eqnarray}
In consequence, the Hessian matrix has the form:
\begin{eqnarray}
  \left[ 
    \begin{array}{cc} 
    N(1,0,\ldots,0) & 0 \\
    0               & H_{jk} \end{array} \right],
\end{eqnarray}
where the $H_{jk}$ are for $j,k = 2,\ldots,d$.  But for such values of
$j$ and $k$ we have $g_{,j} = g_{,k} = 0$ and thus by
Equation~(\ref{eq:Hessian}) we have $H_{jk} = N g_{,jk} / \nabla g
\cdot y$, whence the Hessian matrix has the form:
\begin{eqnarray}
  \left[ 
    \begin{array}{cc} 
    N(1,0,\ldots,0) & 0 \\
    0              & N g_{,jk} / \nabla g \cdot  y 
\end{array} \right].
\end{eqnarray}
The strict positivity of this matrix now follows from our assumption
that the matrix whose entries are the $g_{,jk}$ is strictly positive,
and the fact that any submatrix of a strictly positive matrix is
strictly positive.  \qed

We can sum up the results of the last two propositions in a single
theorem.

\begin{theorem}
  Suppose $g : \mathbb R^d \rightarrow \mathbb R$ is smooth, convex,
  satisfies $g(0) < 1$, is such that $S_g$ is compact, and the matrix
  $G$ with entries $\left( \frac{\partial g}{\partial y^j \partial
      y^k} \right)$ is strictly positive for any $ y \neq 0$ .  Then
  the function $N_g$ defined by the equations
  \begin{eqnarray} 
  N_g(0) & = & 0; \\
  g\left( \frac{ y}{N_g( y)} \right) & = & 1, \mbox{ provided }  y
  \neq 0
  \end{eqnarray}
  exists, is uniquely defined, and is a Minkowski norm with indicatrix
  $S_g$.
\end{theorem}

\subsection{Constructing the approximating Finsler metrics}
\label{subsec:approximations-norm}

We now have all the tools in place to construct the desired Finsler
metrics.  In particular, we now explicitly construct $F_{p\Delta}$.
The family of Finsler metrics $F_{1\Delta}$ follow as the special case
where $p(j) = 1$ for all $j$.

Consider first the function of a single variable $g(y) \equiv
\sqrt{\Delta^2+y^2}$.  This is a smooth and strictly convex function,
but as $\Delta \rightarrow 0$ it approaches $|y|$.  This suggests
defining ($y$ is now a $(4^n-1)$-dimensional vector):
\begin{eqnarray}
  g_{p\Delta}( y) \equiv \sum_\sigma p(\mbox{wt}(\sigma)) 
  \sqrt{\Delta^2+(y^\sigma)^2}.
\end{eqnarray}
Provided $\Delta$ is sufficiently small, it is easy to verify that
$g_{p\Delta}$ is smooth, convex, satisfies $g(0) < 1$, and is such
that $S_g$ is compact.  A calculation shows that
\begin{eqnarray}
  \frac{\partial g_{p\Delta}}{\partial y^\sigma \partial y^\tau} = 
  \frac{p(\mbox{wt}(\sigma) \Delta^2 \delta_{\sigma \tau}}
  {(\Delta^2+(y^\sigma)^2)^{3/2}},
\end{eqnarray}
which clearly specifies a strictly positive matrix.  Thus
$g_{p\Delta}$ induces a Minkowski norm $N_{p\Delta} \equiv
N_{g_{p\Delta}}$.

It is intuitively clear that $N_{p\Delta}$ tends to the norm $N_p(y)
\equiv \sum_\sigma p(\mbox{wt}(\sigma)) |y^\sigma|$ as $\Delta$ goes
to zero.  We can make this intuition quantitative as follows.  Define
$P \equiv \sum_\sigma p(\mbox{wt}(\sigma))$ and observe that
\begin{eqnarray} 
  N_p(y) \leq g_{p\Delta}( y) \leq N_p( y)+P\Delta,
\end{eqnarray}
where the first inequality follows from the fact that $|y| \leq
\sqrt{\Delta^2+y^2}$, and the second inequality follows from the fact
that $\sqrt{\Delta^2+y^2} \leq |y|+\Delta$.  These inequalities imply:
\begin{eqnarray} 
  N_p\left( \frac{ y}{N_{p\Delta}( y)} \right) \leq 
  g_{p\Delta} \left(\frac{ y}{N_{p\Delta}( y)}\right) & \leq &
  N_p\left( \frac{ y}{N_{p\Delta}( y)} \right) \nonumber \\
  & & + P \Delta. 
\end{eqnarray}
Observing that $g_{p\Delta}( y / N_{p\Delta}( y)) = 1$,
multiplying by $N_{p\Delta}( y)$, and rearranging gives
\begin{eqnarray} 
  N_p( y) \leq N_{p\Delta}( y) \leq \frac{N_p(y)}
  {1-P\Delta}.
\end{eqnarray}
Thus provided $\Delta \ll 1/P$ we see that $N_p( y) \approx
N_{p\Delta}( y)$.

Summing up, we have defined $g_{p\Delta} : \mathbb R^{4^n-1}
\rightarrow \mathbb R$ by $g_{p\Delta}( y) \equiv \sum_\sigma
p(\mbox{wt}(\sigma)) \sqrt{\Delta^2 + (y^\sigma)^2}$.  Provided
$\Delta < 1/P$, where $P \equiv \sum_\sigma p(\mbox{wt}(\sigma))$,
there exists a unique function $N_{p\Delta} : \mathbb R^{4^n-1}
\rightarrow \mathbb R$ defined by $N_{p\Delta}(0) \equiv 0$ and for
all other $ y$ by
\begin{eqnarray}
  g_{p\Delta}\left( \frac{ y}{N_{p\Delta}( y)} \right) = 1.
\end{eqnarray}
This function $N_{p\Delta}$ is a Minkowski norm which approximates
$N_p$ well in the sense that
\begin{eqnarray}
  N_p( y) \leq N_{p\Delta}( y) \leq \frac{N_p( y)}{1-P\Delta}.
\end{eqnarray}

We use $N_{p\Delta}$ to define a Finsler metric $F_{p\Delta}$ on
$SU(2^n)$, via
\begin{eqnarray}
  F_{p\Delta}(U,y) \equiv N_{p\Delta}(\gamma),
\end{eqnarray}
where $\gamma$ is the $(4^n-1)$-dimensional vector whose components
are the natural $U$-adapted co-ordinates for $y \in T_U SU(2^n)$.  By
construction this is a family of Minkowski norms on $SU(2^n)$.  To see
that $F_{p\Delta}$ is Finsler, we need only prove that it is a smooth
function of $U$.  This is intuitively clear.  A rigorous proof follows
from the results of Section~\ref{sec:geodesic_equation}, which show
how to explicitly calculate $F_{p\Delta}$.  Note that $F_{p\Delta}$
approximates $F_p$ well in the sense that:
\begin{eqnarray} \label{eq:approximation-result}
  F_p(U, y) \leq F_{p\Delta}(U, y) \leq 
  \frac{F_p(U, y)}{1-P\Delta}.
\end{eqnarray}

%
%
Using results of Ghomi~\cite{Ghomi04a} on smoothing of convex
polytopes it is possible to extend the approximation described here to
show that any right-invariant local metric can be approximated
arbitrarily well by a right-invariant Finsler metric.  Ghomi's results
even imply that any symmetries in the local metric can be retained by
the approximating Finsler metric.  I expect that it is possible to
extend Ghomi's results to approximation of arbitrary local metrics by
Finsler metrics, but have not verified this assertion.  The main
disadvantage of Ghomi's constructions --- a very substantial
disadvantage from our point of view, and the reason they are not used
in this paper --- is that they involve substantially more complex
computations than in the approach we have used to approximate $F_1$
and $F_p$.

\section{Proof of Theorem~\ref{thm:change-qubit-coords}}
\label{app:qubit-change-vars}

To prove Theorem~\ref{thm:change-qubit-coords} we make use of the
following lemma:

\begin{lemma} {} \label{lemma:co-ord-change}
  The equation $\vec X + \vec X \times \vec A = \vec B$, where $\vec
  A, \vec B$ and $\vec X$ are all three-dimensional real vectors, has
  the unique solution
  \begin{eqnarray}
     \vec X = \frac{1}{1+\| \vec A\|^2}\left(\vec B + \vec A \, \vec A
       \cdot  \vec B + 
     \vec A \times \vec B\right).
  \end{eqnarray}
\end{lemma}

\textbf{Proof of Lemma~\ref{lemma:co-ord-change}:} This solution is
easily verified by hand or using any of the standard computer algebra
packages. \qed

\textbf{Proof of Theorem~\ref{thm:change-qubit-coords}:} Fixing $\vec
y$ it is clear that some $\tilde y$ satisfying
Equation~(\ref{eq:change-qubit-coords}) must exist.  All that we have
to do is verify that $\tilde y$ has the form specified in
Equation~(\ref{eq:tilde-y-qubit}).  To do this we simply compare the
order $t$ terms on the left- and right-hand sides of
Equation~(\ref{eq:change-qubit-coords}).  Beginning with the
right-hand side we see that the term of order $t$ is:
\begin{eqnarray}
  & & -i t ( \tilde y \cdot \sigma ) 
  \exp(-i \vec x \cdot \sigma) = \nonumber \\
& & -t [ \sin(\| \vec x\|) 
\hat x \cdot \tilde y I +i  \left(
      \cos(\|\vec x\|) \tilde y + \sin(\| \vec x\|) 
      \tilde y \times \hat x \right) \cdot \sigma ]. \nonumber \\
    \label{eq:rhs-term-t}
\end{eqnarray}

To compute the terms of order $t$ obtained from the left-hand side of
Equation~(\ref{eq:change-qubit-coords}) it helps to define $\vec z
\equiv \vec x + t \vec y$.  Simple calculations show that the
following relationships hold, all to first order in $t$:
\begin{eqnarray}
  \| \vec z \| & = & \| \vec x \| +t \hat x \cdot \vec y \\
  \cos(\|\vec z\|) & = & \cos(\|\vec x\|)-t \sin(\|\vec x\|) \hat x \cdot
  \vec y \\
  \sin(\|\vec z\|) & = & \sin(\|\vec x\|)+t \cos(\|\vec x\|) \hat x \cdot
  \vec y \\
  \hat z & = & \hat x + \frac{t}{\|\vec x\|} \vec y_\perp,
\end{eqnarray}
where $\vec y_\perp \equiv \vec y - \hat x \cdot \vec y \, \hat x$ is
the component of $\vec y$ orthogonal to $\vec x$.  Expanding the
left-hand-side of Equation~(\ref{eq:change-qubit-coords}) out gives
\begin{eqnarray}
  & & \cos(\|\vec z\|)I-i \sin(\| \vec z\|) \hat z \cdot \sigma \\
  & = & \left( \cos(\| \vec x\|)-t \sin(\|\vec x\|) \hat x \cdot \vec y 
  \right) I \nonumber \\
  & & -i \left( \sin(\| \vec x\|)+t \cos(\|\vec x\|) \hat x \cdot  \vec y 
  \right)
  \left( \hat x + \frac{t}{\|\vec x\|}  \vec y_\perp \right)\cdot \sigma.
 \nonumber \\
  & & 
\end{eqnarray}
It follows that the term of order $t$ on the left-hand side of
Equation~(\ref{eq:change-qubit-coords}) is
\begin{eqnarray} \label{eq:lhs-term-t}
  & -t [& \!\! \sin(\|\vec x\|) \hat x \cdot \vec y \, I \nonumber \\
  & & +i\left(
      \cos(\|\vec x\|) \hat x \cdot \vec y \, \hat x + \sinc(\|\vec x\|) 
      \vec y_\perp \right) \cdot \sigma ].
\end{eqnarray}
Comparing the terms in Equations~(\ref{eq:rhs-term-t})
and~(\ref{eq:lhs-term-t}) we obtain two equations:
\begin{eqnarray} 
  \label{eq:condition-1}
   \vec x \cdot \vec y & = & \vec x \cdot \tilde y  \\
  \label{eq:condition-2}
  & & \cos(\| \vec x\|) \hat x \cdot \vec y \, \hat x + 
  \sinc(\| \vec x\|) \vec y_\perp \nonumber \\
  & = &
  \cos(\|\vec x\|) \tilde y + \sin(\|\vec x\|) \tilde y \times \hat x. 
\end{eqnarray}
We will use these equations to solve for $\vec y$ in terms of $\tilde
y$, and vice versa.  Let us first express $\vec y$ in terms of $\tilde
y$.  To do this it helps to note that $\vec y_\perp =\vec y - \hat x
\cdot \vec y \, \hat x = \vec y - \hat x \cdot \tilde y \, \hat x$, by
Equation~(\ref{eq:condition-1}).  Substituting this expression for $
\vec y_\perp$ into Equation~(\ref{eq:condition-2}), multiplying by
$1/\sinc(\|\vec x\|)$ and simplifying we obtain
\begin{eqnarray} 
   \vec y = \tilde y_{\|} + \| \vec x \| \cot(\|\vec x\|) 
   \tilde y_\perp + \tilde y \times \vec x,
\end{eqnarray}
where $\tilde y_{\|} \equiv \hat x \cdot \tilde y \, \hat x$ is the
component of $\tilde y$ parallel to $\hat x$, and $\tilde y_\perp
\equiv \tilde y - \tilde y_{\|}$ is the component of $\tilde y$
orthogonal to $\hat x$.  This is the desired expression for $\vec y$
in terms of $\vec x$ and $\tilde y$.

To obtain $\tilde y$ in terms of $\vec y$ and $\vec x$, we again start
from Equation~(\ref{eq:condition-1}) and~(\ref{eq:condition-2}).
Multiplying Equation~(\ref{eq:condition-2}) by $1/\cos(\|\vec x\|)$ we
see it is equivalent to
\begin{eqnarray}
  \tilde y + \tan(\|\vec x\|) \tilde y \times \hat x = \hat x \cdot \vec y 
  \, \hat x
  +\frac{\tan(\| \vec x\|)}{\|\vec x\|}  \vec y_\perp.
\end{eqnarray}
Applying Lemma~\ref{lemma:co-ord-change} and simplifying the resulting
expression we obtain
\begin{eqnarray}
  \tilde y =  \vec y_{\|} + \sinc(2\|\vec x\|)  \vec y_\perp +
  \sinc^2(\|\vec x\|)  \vec x
  \times  \vec y_\perp,
\end{eqnarray}
where $\vec y_{\|} \equiv \hat x \cdot \vec y \, \hat x$ is the
component of $\vec y$ in the $\hat x$ direction.  This is the desired
expression for $\tilde y$ in terms of $\vec x$ and $\vec y$.

\qed

\section{Vectorizing matrix equations}
\label{app:vectorizing}

In this appendix we give a brief introduction to the
\emph{vectorizing} technique, which can be used to convert matrix
equations into equivalent vector equations.  The treatment is based
on~\cite{Nielsen04d}, which is, in turn, based on material in
Chapter~4 of Horn and Johnson~\cite{Horn91a}.

The vectorizing technique is based on a mathematical operation known
as the \emph{vec} operation, which may be applied to either a matrix
or a superoperator.  When $\mbox{vec}$ is applied to a matrix, it
produces as output the vector formed by stacking all the columns of
the matrix up on top of one another.  More formally, let $M_{m,n}$
denote the space of $m \times n$ complex matrices.  Let $A \in
M_{m,n}$.  Then we define $\mbox{vec}(A)$ to be the $mn$-dimensional
vector formed by stacking all the columns of $A$ up on top of one
another.  For example, we have:
\begin{eqnarray}
  A = \left[ \begin{array}{cc} a & b \\ c & d \end{array} \right]
  \Rightarrow \mbox{vec}(A) = \left[ \begin{array}{c} a \\ b \\ c \\ d
      \end{array} \right].
\end{eqnarray}
We call $\mbox{vec}(A)$ the \emph{vectorized} form of the matrix $A$.

The operation $\mbox{vec}$ has many useful properties, and we note
only a few here (see~\cite{Nielsen04d,Horn91a} for more).  In
particular, if $A,B \in M_{m,n}$ then $\mbox{vec}(A) = \mbox{vec}(B)$
if and only if $A = B$.  Furthermore, for every $mn$-dimensional
vector $v$ there exists a unique matrix $M \in M_{m,n}$ such that
$\mbox{vec}(M) = v$.  We will write $M = \mbox{unvec}(v)$, and speak
of \emph{unvectorizing} $v$ to obtain $M$.  Note that for this
operation to be well-defined we need to specify $m$ and $n$.

%
%
Why define $\mbox{vec}$?  The answer is that it provides an
algebraically and computationally convenient way of making explicit
the structure of $M_{m,n}$ as a vector space.

%
%
The key algebraic fact about $\mbox{vec}$ can be understood physically
as a connection with maximally entangled states.  Let $A \in M_{m,n}$,
and let quantum systems $Q_1$ and $Q_2$ both have dimension $n$.
Define an (unnormalized) maximally entangled state of $Q_1 Q_2$ by
\begin{eqnarray}
  |ME_n\rangle \equiv \sum_j |j\rangle |j\rangle
\end{eqnarray}
where the $|j\rangle$ are fixed orthonormal bases for systems $Q_1$
and $Q_2$, respectively. (We won't bother to distinguish the two bases
notationally, although they are, of course, distinct bases.)
Regarding the matrix $A$ as being defined in the basis $|j\rangle$ for
$Q_2$, we have the identity
\begin{eqnarray}
  \label{eq:vec-identity} \mbox{vec}(A) = (I_n \otimes A)
  |ME_n\rangle,
\end{eqnarray}
where $I_n$ is the $n \times n$ identity matrix.  We will omit the
subscript $n$ when its value is clear from context.  To prove
Equation~(\ref{eq:vec-identity}) note that by linearity it suffices to
prove the identity when $A = |j\rangle \langle k|$.  The proof is
completed by verifying that $\mbox{vec}(|j\rangle \langle k|) =
|k\rangle |j\rangle$ and $(I \otimes |j\rangle \langle
k|)|ME_n\rangle = |k\rangle |j\rangle$.

The identity Equation~(\ref{eq:vec-identity}) has an extremely useful
generalization, which Horn and Johnson ascribe to Roth~\cite{Roth34a}.
The proof is straightforward algebraic manipulation, and thus is
omitted.

\begin{lemma}[Roth's lemma] {} \label{lemma:Roth}
  When $A \in M_{l,m}, B \in M_{m,n}, C \in M_{n,p}$, we have
  \begin{eqnarray}
    \mbox{vec}(ABC) = (C^T \otimes A) \mbox{vec}(B).
  \end{eqnarray}
\end{lemma}

Roth's lemma is extremely helpful in the analysis of linear matrix
equations, such as $\sum_{j} A_j X B_j = C$.  From Roth's lemma, we
see that this is equivalent to the equation
\begin{eqnarray}
  \sum_j (B_j^T \otimes A_j) \mbox{vec}(X) = \mbox{vec}(C),
\end{eqnarray}
which may be solved using standard techniques.  

This discussion suggests defining a vectorized form for a
superoperator.  In particular, given a superoperator ${\cal
  L}(\cdot)$, we define a vectorized form of ${\cal L}$ as follows.
First, note that ${\cal L}$ can always be written in the form ${\cal
  L}(X) = \sum_{j} A_j X B_j$, for some set of matrices $A_j$ and
$B_j$.  Then the vectorized form of ${\cal L}$ is defined by
\begin{eqnarray}
  \mbox{vec}({\cal L}) \equiv \sum_j B_j^T \otimes A_j.
\end{eqnarray}
It is not difficult to show that $\mbox{vec}({\cal L})$ defined in
this way is unique, i.e., it does not depend on the particular
representation in terms of a set of $A_j$ and $B_j$ operators.  By
Roth's lemma we have
\begin{eqnarray}
  \mbox{vec}({\cal L}) \mbox{vec}(X) = \mbox{vec}( {\cal L}(X)).
\end{eqnarray}

With these definitions we see that the vectorized forms of the
superoperators ${\cal I}$ and $\ad_X$ are given by
\begin{eqnarray}
  \mbox{vec}({\cal I}) & = & I \otimes I \\ \label{eq:vec-ad}
  \mbox{vec}(\ad_X) & = & I \otimes X - X^* \otimes I,
\end{eqnarray}
where in the second line we assumed that $X$ is Hermitian, and thus
$X^T = X^*$.

The $\mbox{vec}$ operation for superoperators has all the algebraic
properties one would expect.  It is linear in ${\cal L}$, and a
homomorpishm, i.e., $\mbox{vec}({\cal L}_1 \circ {\cal L}_2) =
\mbox{vec}({\cal L}_1)\mbox{vec}({\cal L}_2)$.  That is, $\mbox{vec}$
converts composition of linear superoperators into matrix
multiplication.  As a consequence, we see that if $f(x) =
\sum_{j=0}^\infty f_j x^j$, then $\mbox{vec}(f({\cal L})) =
f(\mbox{vec}(L))$.  Using this result, followed by
Equation~(\ref{eq:vec-ad}), we deduce that
\begin{eqnarray}
\mbox{vec}(\exp(-i \ad_X)) = U^* \otimes U,
\end{eqnarray}
where $U = \exp(-iX)$.  It follows that the operation ${\cal E}_X$
defined in Subsection~\ref{subsec:su2n} has vectorized form:
\begin{eqnarray} \label{eq:app-vec-Ex}
  \mbox{vec}({\cal E}_X) = \frac{U^* \otimes U-I \otimes I}
  {-i(I \otimes X-X^* \otimes I)}.
\end{eqnarray}

\acknowledgments

Thanks to Scott Aaronson, Dorit Aharonov, Michael Ben-Or, Harry
Buhrman, Jennifer Dodd, Andrew Doherty, Mark Dowling, Alexei
Gilchrist, Henry Haselgrove, Matt James, Greg Kuperberg, Nick
Menicucci, David Poulin, Oded Regev, and Mohan Sarovar for stimulating
conversations.


\end{document}